\documentclass[%
 reprint,
 amsmath,amssymb,
 aps,
floatfix
]{revtex4-1}

\usepackage{graphicx}
\usepackage{dcolumn}
\usepackage{bm}
\usepackage[colorlinks=true]{hyperref}
\usepackage{siunitx}
\usepackage{subfigure}
\usepackage[T1]{fontenc}
\usepackage{float}

\usepackage{xcolor}
\usepackage{soul}
\usepackage{systeme}

\begin{document}
\newtheorem{THR}{Theorem}
\newtheorem{DEF}{Definition}
\newtheorem{LEM}{Lemma}
\newtheorem{COL}{Corollary}
\newtheorem{EXMP}{Example}
\newtheorem{PROB}{Problem}
\newtheorem{ALG}{Algorithm}
\newtheorem{PROP}{Proposition}
\newtheorem{REM}{Remark}

\newcommand*\Laplace{\mathop{}\!\mathbin\bigtriangleup}
\newcommand{\Tau}{\mathcal{T}}

\def\BT{\begin{THR}} 	        \def\ET{\end{THR}} 
\def\BL{\begin{LEM}} 	        \def\EL{\end{LEM}} 
\def\BD{\begin{DEF}} 	        \def\ED{\end{DEF}}
\def\BC{\begin{COL}} 	        \def\EC{\end{COL}} 
\def\BE{\begin{equation}}		\def\EE{\end{equation}}
\def\BSE{\begin{subequations}}  \def\ESE{\end{subequations}}
\def\BEA{\begin{eqnarray}}		\def\EEA{\end{eqnarray}}
\def\BG{\begin{gather}}			\def\EG{\end{gather}} 
\def\BM{\begin{matrix}}			\def\EM{\end{matrix}} 
\def\BA{\begin{array}}			\def\EA{\end{array}} 
\def\BAn{\begin{align}}         \def\EAn{\end{align}}

\def\BS{\begin{split}}			\def\ES{\end{split}} 
\def\BP{\proof}					\def\EP{$\square$} 
\def\BEX{\begin{EXMP}}			\def\EEX{\end{EXMP}} 
\def\BPR{\begin{PROB}}			\def\EPR{\end{PROB}} 
\def\BAl{\begin{ALG}}			\def\EAl{\end{ALG}} 
\def\BProp{\begin{PROP}}			\def\EProp{\end{PROP}} 
\def\BRemark{\begin{REM}}            \def\ERemark{\end{REM}}

\def\R{$\mathbb{R}$ }
\def\C{$\mathbb{C}$ }
\def\N{$\mathbb{N}$ } 
\def\Q{$\mathbb{Q}$ } 
\def\Z{$\mathbb{Z}$ } 
\def\E{$\mathbb{E}$ } 
\def\E{$\mathbb{W}$ }

\def\Rr{\mathbb{R} } 
\def\Cc{\mathbb{C} }
\def\Nn{\mathbb{N} }
\def\Qq{\mathbb{Q} }
\def\Zz{\mathbb{Z} }
\def\Ee{\mathbb{E} }
\def\Ee{\mathbb{W} }

\def\Xx{\mathbf{X} }
\def\Yy{\mathbf{Y} }
\def\Mm{\mathbf{Mm} }
\def\Iy{\infty}

\def\RM{$\Leftrightarrow$ } 
\def\HENCE{$\Rightarrow$ } 
\newcommand{\norm}[1]{\left\lVert#1\right\rVert}


\preprint{APS/123-QED}

\title{Mechanism of overscreening breakdown by electrode surface morphology in asymmetric ionic liquids}
%
\author{Irina Nesterova}%

\author{Aleksey Khlyupin}
\email{khlyupin@phystech.edu}
\affiliation{%
 Moscow Institute of Physics and Technology, Institutskiy Pereulok 9, Dolgoprudny, Moscow, Russia, 141700
}%


\author{Nikolay Evstigneev}
\author{Oleg Ryabkov}
\affiliation{
 Federal Research Center “Computer Science and Control”,
Institute for System Analysis, Russian Academy of Science, Vavilova str., 40, Moscow, Russia, 119333
}%

\author{Kirill M. Gerke}
\affiliation{%
 Schmidt Institute of Physics of the Earth of Russian Academy of Sciences, Bolshaya Gruzinskaya 10, Moscow, Russia, 123242
}%


\date{\today}

\begin{abstract}

The interfacial nature of the electric double layer (EDL) assumes that electrode surface morphology significantly impacts the EDL properties. Since molecular-scale roughness modifies the structure of EDL, it is expected to disturb the overscreening effect and alter differential capacitance (DC). In this Letter, we present a model that describes EDL near atomically rough electrodes with account for short-range electrostatic correlations. We provide numerical and analytical solutions for the analysis of conditions for the overscreening breakdown and DC shift estimation. Our findings reveal that electrode surface structure leads to DC decrease and can both brake or enhance overscreening depending on the relation of surface roughness to electrostatic correlation length and ion size asymmetry.

\end{abstract}

\maketitle

\section{\label{sec:introduction}Introduction}

The structure of EDL is an important fundamental characteristic that plays a key role in applications such as energy storage, colloidal physics, biophysics, and electrocatalysis technologies \cite{schellman1977electrical, henderson2009insights, gur2018review, shin2022importance}. The classical model proposed by Gouy and Chapman \cite{gouy1910constitution, chapman1913li} represents the EDL structure as a large diffuse layer near the electrode surface, where the predominant concentration of counterions compensates the charge on the electrode, but this model disregards the finite ion size constraints. Later, the contribution of ion sizes was considered, for instance, by using continuum volume constraints \cite{bikerman1942xxxix}, or with the application of mean-field theory \cite{kornyshev2007double}. Although the expression for charge density distribution was reformulated, still there was no representation of ions layering near the electrode surface. After, Bazant, Storey and Kornyshev (BSK) introduced the model accounting for short-range electrostatic correlations \cite{bazant2011double}. Their model provided significant modifications in the EDL structure, including charge density oscillations, overscreening, and crowding regimes. However, in all these works, EDL was considered near a flat electrode surface, while real electrode surfaces have molecular-scale heterogeneous surface structures \cite{shi2002scanning, park2009immobilization}.

The morphological characteristics of the electrode surface change the response of EDL to surface charge and determine double layer properties. Previous studies proved that real surfaces with molecular scale surface roughness can disrupt the layering structure \cite{sheehan2016layering, neimark2009quenched, aslyamov2017density, pean2015single, simon2020perspectives}.
Accordingly, surface roughness is suggested to modify the overscreening effect \cite{pean2015single, goodwin2017mean, aslyamov2021electrolyte, aslyamov2022properties}. In this regard, the impact of electrode surface geometry on EDL structure was investigated in MD studies \cite{vatamanu2011influence, vatamanu2012molecular}. They demonstrate that rough electrode surfaces can decline, shift, and spread ion density distribution peaks. Furthermore, the evidence of overscreening breakdown by electrode structure was obtained by MD simulations on porous material \cite{merlet2012molecular}. 

\begin{figure}
\centering
\includegraphics[width=1\linewidth]{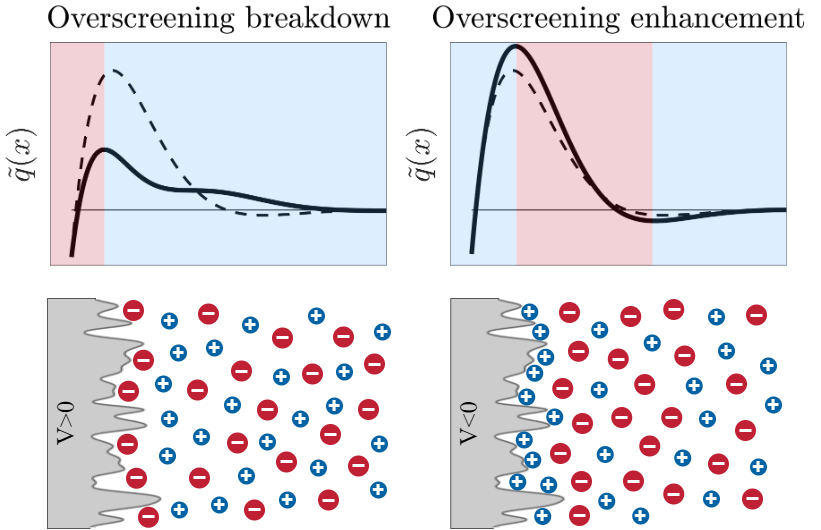}
\caption{The scheme of ion separation for ionic liquid with bigger anions near the rough electrode surface, where the scale of roughness is smaller than electrostatic correlation length, and corresponding normed cumulative charge $\tilde q(x)$ for the cases: overscreening breakdown at $V>0$ (on the left), and overscreening enhancement at $V<0$ (on the right).}
\label{fig:toc}
\end{figure}

To account for molecular-scale surface structure, Aslyamov et al. proposed a mean-field model for EDL near rough electrodes \cite{aslyamov2021electrolyte}. The authors derived an expression for the ion charge density with an account for electrode surface roughness and numerically examined the impact of surface structure on DC. Later, Khlyupin et al. developed an analytical model accounting for ion separation in asymmetric ionic liquids near rough electrode surface \cite{khlyupin2023molecular}. They have demonstrated DC behaviour transition due to surface roughness and successfully captured the third peak in the DC profile caused by ions reorientation. However, these models lack electrostatic correlations.

This Letter introduces a model that combines the impact of surface morphology \cite{khlyupin2023molecular} and short-range electrostatic correlations \cite{bazant2011double}. The interplay between electrode surface roughness and electrostatic correlation is investigated numerically. Furthermore, we provide analytical solutions for the obtained Modified Poisson--Fermi equation near rough electrodes together with the shift of DC. Quite importantly, the model describes two regimes when the scale of surface roughness is smaller or greater than the electrostatic correlation length.  At small roughness scales, it creates the additional ion-specific force that either enhances or disrupts overscreening, depending on ion size asymmetry or the electrode charge, as schematically shown in Fig.\ref{fig:toc}. At larger scales, roughness acts like chemical potential, keeping the structural inhomogeneity of the EDL, but reducing the cumulative charge. The latter effect can also strengthen the overscreening effect or make it weaker, according to the charge on the electrode and ion sizes.

\section{Methods}

\subsection{Theoretical model with electrostatic correlations near rough electrode}

To describe electrostatic correlations, Bazant, Storey, and Kornyshev \cite{bazant2011double} extended the formulation for the total free energy $G$ of ionic liquid using the Landau-Ginsburg functional. They considered electrostatic potential perturbation described by a non-local contribution of ion-ion correlations. Keeping the first term of gradient expansion from non-local ion-ion correlations and applying $\partial G/\partial \phi = 0$ they obtained the modified Poisson equation:
\begin{equation}\label{eq:BSK}
    (1-\delta_c \nabla^2)\nabla^2 \phi = -\rho,
\end{equation}
which determine the relation between the electrostatic potential $\phi$ and the charge density function $\rho$ involving dimensionless correlation length of the electrostatic field $\delta_c = l_c/L_D$ normed at the Debye length $ L_D$.

\begin{figure}
\centering
\includegraphics[width=1\linewidth]{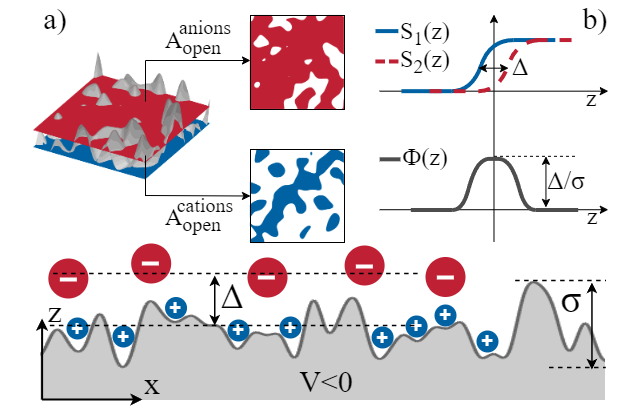}
\caption{Schematic illustration of ion separation near rough electrode surface at $V<0$, where $\Delta$ is the difference of ion penetration depths and $\sigma$ is the height deviation of rough electrode surface. Inset (a): the area permitted for cations and anions $A_{open}$ at corresponding levels. Inset (b): characteristic functions $S(z)$ for cation and anions and their difference $\Phi(z)$ that describes ion separation.}
\label{fig:model_scheme}
\end{figure}

In this study, we incorporate the influence of electrode morphological characteristics by accounting for the constraints on the permitted area and non-local ion-ion interactions caused by electrode surface structure. To account for steric restriction between an ion and electrode surface, we use the characteristic function $S_i(z)$, where index $i$ corresponds to ion type. This function reflects the ratio of the permitted area for the ion on level $z$ away from the surface that is determined by the surface structure and ion size (see Fig.\ref{fig:model_scheme}). 
We utilize the expression for the ion charge density, including the characteristic function $S_i(z)$, obtained in Ref. \cite{aslyamov2021electrolyte}:
\begin{equation}\label{eq:ions_distr}
    \rho_i(z)=eZ_ic_0 S_i(z)\frac{e^{-Z_i\beta e \phi}}{1 + \sum_i \gamma_i \left[e^{-Z_i \beta e \phi} - 1\right]},
\end{equation}
where $e$ is the electrode charge, $Z_i$ is the ion charge, $c_0$ is ions bulk concentration, $\beta = 1/k_B T$, and $\gamma_i = c_0 v_i$ is the compacity parameters with ion volume $v_i$ \cite{kornyshev2007double, aslyamov2021electrolyte}. 

Then, one can obtain the expression for the cumulative charge density near a rough electrode, by expressing $\rho(z) = \rho_1(z)+\rho_2(z)$ and reformulating it through the difference of characteristic functions $\Phi(z) = S_1(z) - S_2(z)$ \cite{suppl_info}. Substituting it into Eq.~(\ref{eq:BSK}), we obtain the Modified Poisson-Fermi equation with the additional terms for electrostatic correlations and electrode surface roughness in dimensionless form:
\begin{equation}\label{eq:problem}
(1-\delta_c^2 \nabla^2) \nabla^2 u = \frac{\sinh(u)}{1 +2\gamma \sinh^2\left(u/2\right)} - \frac{0.5 \Phi(x) e^{u}}{1 +2\gamma \sinh^2\left(u/2\right)},
\end{equation}
where $x = z/L_D$, $u = \beta e \phi$, and we consider $Z_i = \pm1$ and $\gamma = \gamma_1 = \gamma_2$ for simplicity. This equation is solved with the boundary conditions: $u|_{x=0}= V$, where $V$ is dimensionless electrostatic potential on the electrode and $\quad u'''|_{x=0}= 0$, which reflects electroneutrality near the electrode surface. These boundary conditions provide consistent level of the accuracy regarding to the application of BSK model for short-range correlations \cite{suppl_info}.

The second term in the r.h.s. of Eq.~(\ref{eq:problem}) reflects soft repulsion for the anions, because we assume that they have bigger size. The function $\Phi(x) = \mu g(x)$, where $\mu = \Delta/\sigma$ is the magnitude of ion separation, $g(x) = \frac{1}{\sqrt{2 \pi}} \exp{\{-\frac{x^2}{2(\sigma/L_D)^2}\}}$ is monotonically decreasing function with characteristic scale of surface roughness \cite{khlyupin2023molecular}. The surface height deviations $\sigma$ and difference of ion penetration depths $\Delta$ control ion separation of asymmetric ionic liquid near a rough electrode (see Fig.\ref{fig:model_scheme}). The penetration depth depends on the ion size and electrode surface geometry and determines how close the ion approaches the electrode surface. For electrolytes with ion size asymmetry, there will be difference of ion penetration depths, reflecting ion separation performance.

\nocite{shen2009some}
\nocite{Black_1998}
\nocite{Gao_2012}
\nocite{Gottlieb_1977}
\nocite{Shen_2006}
\nocite{Hammad_2020}
\nocite{Boyd_1987}
\nocite{Ortega_1968}
\nocite{Abbasbandy_2007}
\nocite{B_rgisser_2013}
\nocite{studholme1999overlap}

\subsection{Numerical solution}

We calculate the EDL properties \cite{ovrscr_bd_num_solver}, solving Eq.~(\ref{eq:problem}) numerically within a simple 1D problem in the area $x \in [0, +\infty)$. In general, the problem Eq.~(\ref{eq:problem}) can be formulated as:
\begin{equation}
\label{eq:general_problem}
\alpha \frac{\partial^4 u}{\partial x^4} + \beta \frac{\partial^2 u}{\partial x^2} = f(u,x),
u(0) = u_0, \frac{\partial^3 u(0)}{\partial x^3} = 0,
\end{equation}
where $\alpha$, $\beta$ are finite real numbers and $f(u,x)$ is a smooth non-linear function, which is regular at infinity, i.e. $f(u,x)$ and all its derivatives decay to $0$ as $x \to +\infty$ either algebraically or exponentially for regular solutions $u(x)$.

To solve Eq.~(\ref{eq:general_problem}), we apply the pseudospectral collocation method using Chebyshev polynomials in trigonometric form as basis together with analytical mapping between unbounded and bounded domains. Benefits and all technical details are discussed in Ref. \cite{suppl_info}. The discrete nonlinear problem is solved using Newton--Raphson method \cite{ypma1995historical} with globalization that uses homotopy between the provided non-linear function (r.h.s. of Eq.~(\ref{eq:problem})) and a regular exponentially decaying function $\sim \exp(-x)$.

The resulting solution $u(x)$ is expanded in terms of basis functions in the physical domains as follows:
\begin{equation}\label{eq:num_sol}
u(x) = \sum_{j=0}^{\infty} a_j \cos \left[ 2 j\, \text{arccot} \left( \sqrt{x/L} \right) \right]
\end{equation}
here $a_j$ are the expansion coefficients on a set of Chebyshev polynomials $\cos(j t)$ in trigonometric form, with mappings from basis domain $t \in[0,\pi]$ to physical domain $x \in [0, +\infty]$ by the relation: $t=2\text{arccot} \left( \sqrt{x/L} \right)$, where $L$ is a mapping parameter \cite{suppl_info}.

\subsection{Analytical solutions}

Furthermore, we develop analytical solutions for Eq.~(\ref{eq:problem}) in addition to the numerical one. We employ a perturbation theory \cite{khlyupin2023molecular} with a small parameter $\mu \ll 1$ and express the solution as $u = u_0 + \mu u_1$. The first term represents the solution near a flat electrode surface \cite{bazant2011double} and the second term accounts for the perturbation caused by surface roughness. Thus, we derive the following systems to solve for the roughness-induced perturbation at low potentials:
\begin{equation}\label{eq:an_problem}
\left\{
\begin{array}{lll}
\delta_c^2 \nabla^4 u_1 - \nabla^2 u_1 + u_1 = \frac{1}{2} g(x) e^{u_0}\\
    \left.u_1 \right|_{x=0} =0  \\
    \left.u'''_1 \right|_{x=0} =0  \\
\end{array}
\right..
\end{equation}

Below, we provide two analytical solutions of the system (\ref{eq:an_problem}) with different spatial behavior that are switched at $\sigma \approx \delta_c$ \cite{suppl_info}. We compare the results of presented analytical solutions with numerical calculations for various parameters of surface roughness $\sigma$ and find good qualitative and quantitative agreement \cite{suppl_info}.

For small values of surface roughness $\sigma < \delta_c$, we use low potential approximation to simplify r.h.s. of Eq.~(\ref{eq:an_problem}) $e^{u_0} \approx (1+u_0)$ and replace $g(x)$ with a step like function with the same magnitude $1/\sqrt{2\pi}$ for $x \in [0, \tau]$. Then, the solution of Eq.~(\ref{eq:an_problem}) has the form:
\begin{equation} \label{eq:sol_small_sigma}
    u_1 = \frac{1}{2\sqrt{2\pi}}[1 + e^{-kx}(\sin(kx)-\cos(kx))],
\end{equation}
with $k = 1/\sqrt{2 \delta_c}$. The solution is determined for $x<\tau$, where $\tau = 0.5\sigma$ were selected by comparison with the numerical solution. For $x>\tau$, we utilize the general solution of Eq.~(\ref{eq:an_problem}) connected with the solution from Eq.~(\ref{eq:sol_small_sigma}) at $x = \tau$ in the following way: $u_1^I(\tau) = u_1^{II}(\tau)$ and $du_1^I(\tau)/dx = du_1^{II}(\tau)/dx$ \cite{suppl_info}. The resulting solution exhibits non-monotonic behavior with oscillations and it is not correlated with potential in the electrode $V$. The perturbation magnitude grows with an increase of surface roughness $\sigma$ till it reaches the value $2 \pi/ k$. 

At high values of surface roughness $\sigma > \delta_c$, we neglect the fourth-order term with short-range correlation in Eq.~(\ref{eq:an_problem}) \cite{suppl_info}. Then, the solution for the electrostatic potential perturbation can be expressed using the Green function as in Ref.\cite{khlyupin2023molecular}:
\begin{equation}\label{eq:sol_big_sigma}
    u_1(x) = \frac{1}{4} \int_0^{\infty} g(x_0) e^{u_0(x_0)}\left[e^{-|x-x_0|} - e^{-|x+x_0|}\right] d x_0,
\end{equation}
This solution has no oscillations and remains consistently positive. Similar to the previous solution, it increases with surface roughness $\sigma$, but the perturbation magnitude now increases with potential on the electrode $V$.

\section{Results}

\begin{figure}
\centering
\includegraphics[width=1\linewidth]{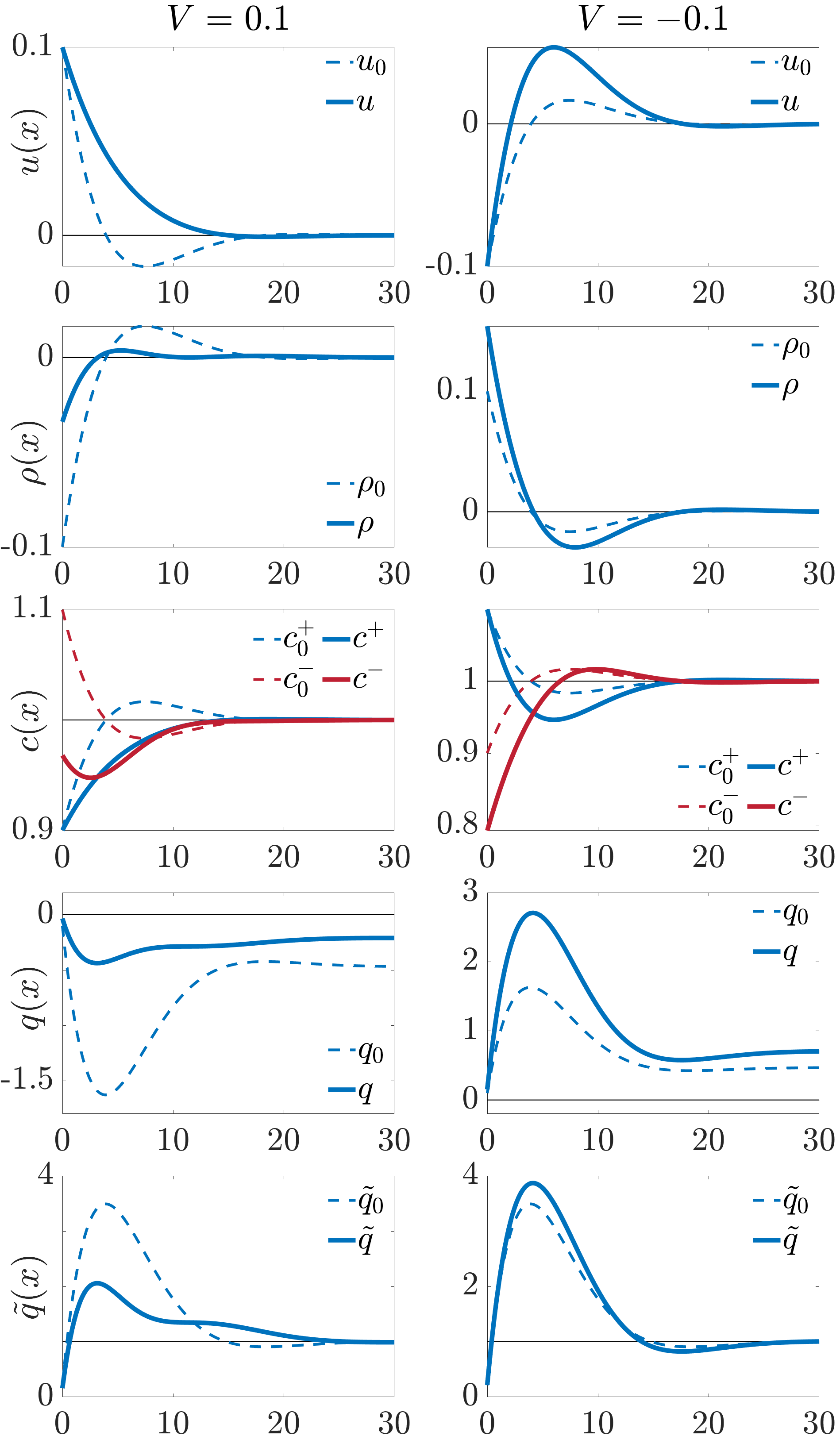}
\caption{Comparison of electrostatic potential $u(x)$, charge density $\rho(x)$, ion concentrations $c_{\pm}(x)$, cumulative charge $q(x)$, and normed cumulative charge $\tilde q(x)$ 
in the EDL near rough (solid lines) and flat (dashed lines) electrode surfaces with ion correlations at $\delta_c = 10, \mu = 0.3$, and $\sigma = 5$. In the left column for $V = 0.1$ and in the right column for $V = -0.1$.}
\label{fig:perturbed}
\end{figure}

To investigate the impact of electrode surface roughness on the EDL structure, we apply the presented numerical model and vary parameters such as the electrode potential $V$, the scale of surface roughness $\sigma$, and the perturbation magnitude $\mu$, keeping $\delta_c = 10$ and $\gamma = 1$. We consider ionic liquid with bigger anions and take $\delta_c = 10$, since it provides better representation with experimental data on DC \cite{bazant2011double}. The EDL near rough electrode is described by electrostatic potential $u(x)$, charge density $\rho(x)$, ion concentrations $c_{\pm}(x)$, cumulative charge of electric double layer $q(x) = \int_0^x \rho(x) dx$, and the normed cumulative charge profile $\tilde q(x) = \int_0^x \rho(x) dx/\int_0^{\infty} \rho(x) dx$. 

\subsection{EDL properties near rough electrode}

Fig. \ref{fig:perturbed} illustrates the main EDL characteristics modified by electrode surface structure. In the study, we take the potential low enough to obtain the overscreening regime near a flat electrode surface. We initially examine the EDL behavior at the positive potential on the electrode $V>0$ with predominant adsorption of anions. If the electrode surface is rough, it provides soft repulsion to anions, so their concentration $c^-(x)$ decreases near the electrode surface, and ion separation become worse. It causes a local increase in charge density $\rho(x)$ and electrostatic potential $u(x)$ near the electrode surface. Consequently, the cumulative charge profile $q(x)$ rises, causing a decrease in the cumulative charge of EDL. As a result, the magnitude of the normed cumulative charge $\tilde q(x)$ dramatically drops and the well of the second layer disappears, reflecting overscreening breakdown.

At negative potential on the electrode $V<0$, when the first layer contain more cations, roughness improves ion separation by repelling anions. It leads to a decrease in the concentration of the anions $c_-(x)$ and an increase in charge density $\rho(x)$ near the electrode. The nonmonotonic behavior of $u(x)$ and $\tilde q(x)$ become more significant. The positive cumulative charge profile $q(x)$ is growing, so the EDL cumulative charge becomes higher. In this case, we observe a slight enhancement of overscreening.

\subsection{Overscreening breakdown and enhancement} 

\begin{figure}
\centering
\includegraphics[width=1\linewidth]{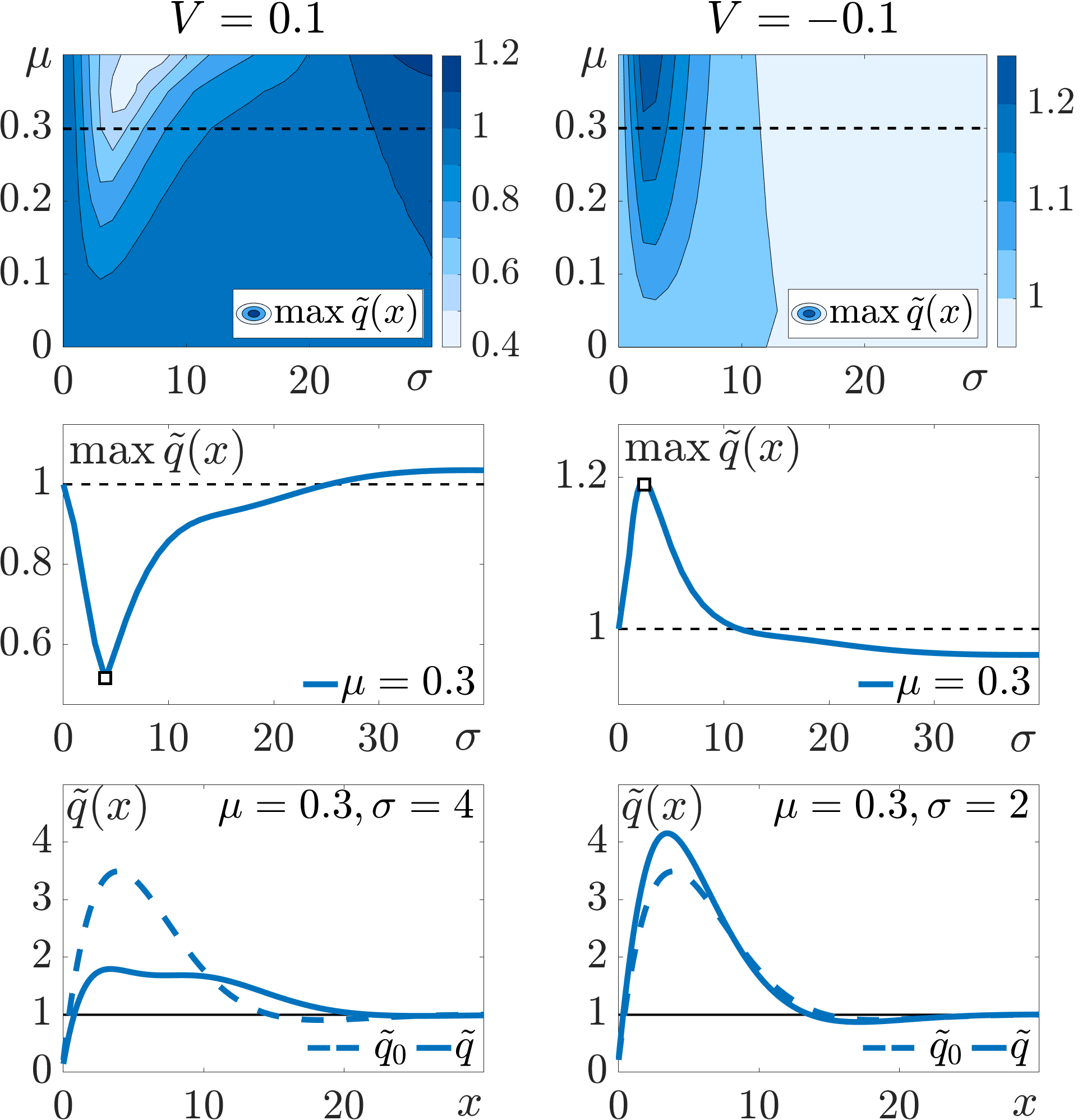}
\caption{The overscreening modifications for $V = 0.1$ in the left column and for $V = -0.1$ in the right column. At the top: maps of the normed cumulative charge $\tilde q(x)$ magnitude depending on the parameters $\mu \in [0, 0.4]$ and $\sigma \in [0, 30]$.  In the center: the profiles of the normed cumulative charge $\tilde q(x)$ magnitude depending on $\sigma \in [0, 30]$ at $\mu = 0.3$. At the bottom: the normed cumulative charge $\tilde q(x)$ profiles at $\mu = 0.3$ and $\sigma = 4$ for $V = 0.1$ or $\sigma = 2$ for $V = -0.1$.}
\label{fig:overscreening_modifications}
\end{figure}

Next, we investigate the behavior of the normed cumulative charge profile $\tilde q(x)$ as it qualitatively and quantitatively describes the overscreening effect. Figure \ref{fig:overscreening_modifications} provides a comprehensive overview of $\tilde q(x)$ modifications depending on the scale of surface roughness $\sigma$ and magnitude of ion separation $\mu$, at $\delta_c = 10$ and for two values of potential $V = 0.1$ (left column) and $V = -0.1$ (right column). At the top, Figure \ref{fig:overscreening_modifications} displays the maps for the normed cumulative charge magnitude $\max \tilde q(x)$ depending on the parameters $\mu$ and $\sigma$. Below, the corresponding $\tilde q(x)$ magnitude line at $\mu = 0.3$ is shown, depending on surface roughness $\sigma$. At the bottom, the normed cumulative charge $\tilde q(x)$ is shown in the cases with the most significant effect of the electrode surface roughness at $\mu = 0.3$ and $\sigma = 4$ for $V = 0.1$ and $\sigma = 2$ for $V = -0.1$.

In the case $V>0$, when $\sigma \le \delta_c$, electrode surface roughness worsens ion separation that leads to a decrease in the $\tilde q(x)$ magnitude and can even destroy overscreening at high $\mu$ values. While, when $\sigma > \delta_c$, the magnitude of $\tilde q(x)$ grows, because rough electrode surface reduces the cumulative charge of EDL more than it modifies the layering that results in excess adsorption of anions for electroneutrality. Conversely, when $V<0$, for $\sigma \le \delta_c$, electrode surface roughness improves ion separation and increases the magnitude of $\tilde q(x)$, enhancing overscreening. However, when $\sigma > \delta_c$, the magnitude of $\tilde q(x)$ decreases, because the rough electrode surface raises the cumulative charge and keeps the structure of EDL, so ion separation becomes less significant. These effects become more explicit at higher values of $\mu$.

Overall, we obtain that the behavior of the normed cumulative charge $ \tilde q(x)$ varies depending on the relation of scales of surface roughness $\sigma$ and electrostatic correlations $\delta_c$. When $\sigma \le \delta_c$, the perturbation significantly modifies EDL structure and can be treated as an ion-specific force. While, when $\sigma > \delta_c$, the rough electrode surface acts more like an additional chemical potential for one type of ion, altering the cumulative characteristics, while keeping the EDL structure. 

\subsection{Value of critical potential}

If the potential on the electrode is high enough, it becomes more challenging to modify the EDL properties by molecular scale surface roughness. At positive potential, for bigger anions, and, when the scale of electrode surface roughness is comparable with electrostatic correlations, electrode morphology creates additional inverse correlation that disrupts the overscreening regime. Important to note, that if we consider negative potential, roughness can not break overscreening. Therefore, we now consider at what critical value of $V^*>0$ electrode surface structure stops to disrupt the overscreening effect. 

The most significant impact on the overscreening is obtained at $\mu = 0.38$ and $\sigma = 6$ for $V = 0.1$ (see the upper left plot in Fig. \ref{fig:overscreening_modifications}) and this effect remains consistent for other positive values of potential on the electrode. Then, we consider the Shannon—Wiener entropy to estimate the EDL structural modifications near the rough electrode surface with selected $\mu$ and $\sigma$ values \cite{suppl_info}. The critical potential value $V^* = 0.24$ is defined at the peak of the entropy gradient with respect to $V$. It means that for $V\ge 0.24$, there are no substantial modifications in the EDL structure caused by electrode roughness.

\subsection{Differential capacitance}

\begin{figure}[t]
\centering
\includegraphics[width=0.85\linewidth]{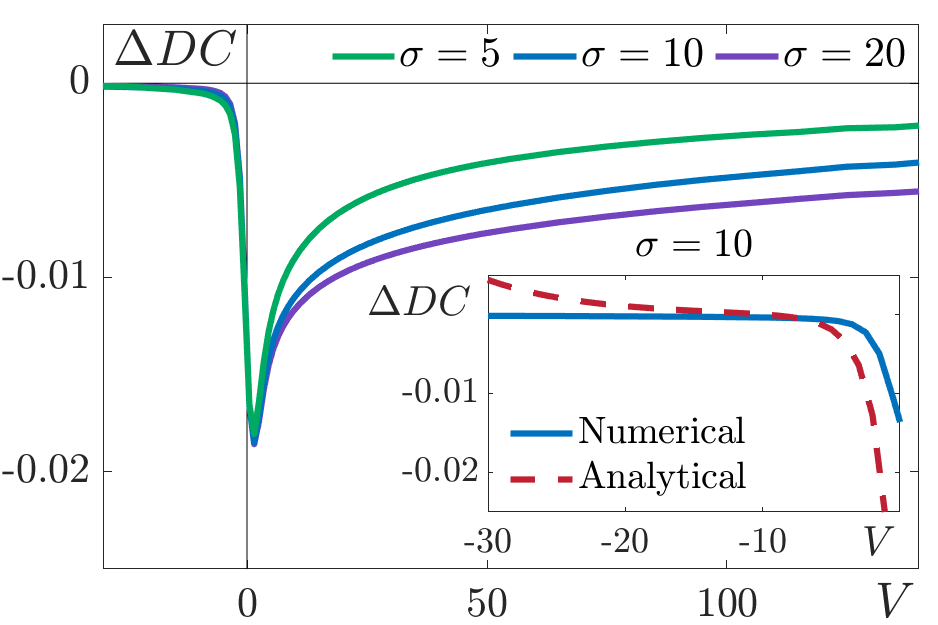}
\caption{DC decrement induced by electrode surface roughness calculated by our numerical model for $\delta_c = 10, \gamma = 0.5$, and $\mu = 0.4$ for various values of $\sigma = 5, 10$, and $20$. In the inset: DC decrement calculated numerically (solid line) and analytically (dashed line) for $\sigma = 10$.}
\label{fig:dc_sigma}
\end{figure}

In addition, we calculate the differential capacitance (DC) of EDL taking into account the influence of electrode surface structure. Fig.\ref{fig:dc_sigma} shows that electrode roughness decreases DC, especially at low positive potentials. It is noteworthy that at high negative potentials, there is no impact on DC since there are no anions to repel near the electrode surface.  Besides, we observe that DC decrement is growing with an increase of the scale of electrode surface roughness $\sigma$ . For convenience, we derive the analytical expression for DC decrement, using $\Delta DC = -\frac{\partial}{\partial V} \frac{\partial \mu u_1(x)}{\partial x}$ and the solution for $u_1$ from  Eq.~(\ref{eq:sol_big_sigma}) \cite{suppl_info}. Applying the following assumptions: $V \ll 1$ and $\sigma > \delta_c \gg 1$, where $k \approx 1/\sqrt{2\delta_c}$, we obtain a constant value for DC decrement near $V = 0$:
\begin{equation}
    \Delta DC \approx -\frac{1}{2 \sqrt{2\pi}}\frac{\Delta}{\sigma} \frac{2k + 1}{2k^2+2k + 1},
\end{equation} 
Thus, soft repulsion of anions worsens ion separation and decreases the differential capacitance of EDL. 

\section{Conclusion}

This study investigated the impact of electrode surface roughness on the EDL structure. Based on the BSK theory, we developed a model for the description of electrode structure impact on EDL properties. We provided numerical \cite{ovrscr_bd_num_solver} and analytical solutions that describe the behavior of EDL near a rough electrode surface. Our findings reveal that electrode surface structure can either enhance or disrupt the overscreening effect. These effects depend on the electrode charge, ion size asymmetry, and the scales of surface roughness and electrostatic correlations. Additionally, we found that DC reduces near rough electrode surfaces. It is important to note that described effects are the especially pronounced at low potentials.

\begin{acknowledgments}
The authors appreciate the partial support from the grant by Russian Science Foundation (RSF) project number 23-21-00095.
\end{acknowledgments}
\bibliography{references}
\bibliographystyle{apsrev}
\newpage
\widetext

\appendix

\def\tocname{Supplementary Information}
\maketitle

\section{Theoretical model}

\subsection{Charge density near rough electrode}

\begin{figure}
\centering
\includegraphics[width=0.9\linewidth]{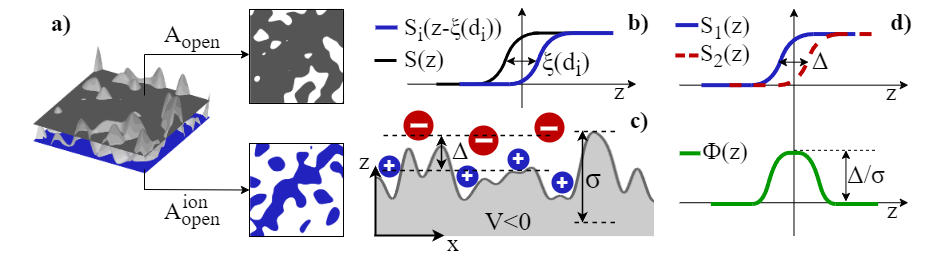}
\caption{(a)~--- Schematic illustration of the available area near the electrode surface (grey) and the shifted available area for the ion with finite size (blue), (b)~--- Characteristic functions of the effective solid potential $S(z)$ and the shifted function on the value $\xi(d_i)$ for the ion with size $d_i$, (c)~--- Schematic illustration of ion separation near the rough electrode surface, (d)~--- Characteristic functions of the effective solid potential for cation $S_1(z)$ and anion $S_2(z)$ with different sizes, where cation is smaller than anion, and the $\Phi(z)$ formed by their difference.}
\label{fig:charge_density_perturbation}
\end{figure}

In this section, we derive the expression for the charge density near a rough electrode surface that we use in the Modified Poisson-Fermi equation (see Eq.(3) in the main text). The influence of rough electrode surface is described by effective solid potential $U^{eff}(z) = -\frac{1}{\beta} \log S(z)$, where $S(z)$ is the characteristic function of effective solid potential given by: $S(z) = A_{open}(z)/(A_{open}(z) + A_{solid}(z))$. It reflects the fraction of the permitted area for ion particles and depends on the geometry of the solid surface. Due to the finite size of ions, they have smaller permitted area, so the characteristic function $S(z)$ is shifted on a value $\xi(d_i)$, which depends on ion diameter (see Fig.\ref{fig:charge_density_perturbation}(a,b)).  The ion density distribution with account for electrode surface geometry was derived based on the asymmetric lattice-gas model in Ref.[18] and include the characteristic function $S_i(z)$:
\begin{equation}\label{eq:ions_distr_0}
\rho_i(z)=eZ_ic_0 S_i(z)\frac{e^{-Z_i\beta e \phi(z)}}{1 + \sum_i \gamma_i \left[e^{-Z_i \beta e \phi(z)} - 1\right]}
\end{equation}
where index $i$ relates to an ion type, $e$ is the charge of an electron, $Z_i$ is the ion charge, $c_0$ is the bulk ion concentration, $\beta = 1/k_BT$, $\phi$ is the electrostatic potential, $\gamma_i = v_ic_0$ is the ion compacity parameter with ion volume $v_i$. 

Further, for simplicity, we use $Z_i = \pm1$, and $\gamma = \gamma_1 = \gamma_2$, where $i = 1,2$  for cations and anions, respectively. Then, the charge density distributions for cations and anions take the following form:
\begin{align}
\label{eq:ions_distr_1}
\rho_1(z)=ec_0 S_1(z)\frac{e^{-\beta e \phi}}{1 +2\gamma \sinh^2\left(\beta e \phi/2\right)} \\
\rho_2(z)=-ec_0 S_2(z)\frac{e^{\beta e \phi}}{1 +2\gamma \sinh^2\left(\beta e \phi/2\right)}
\end{align}

For ionic liquids with different sizes of cations and anions, their characteristic functions $S_i(z)$ are shifted on a value $\Delta$ shown in Fig.\ref{fig:charge_density_perturbation}(c,d), where $\Delta$ is the difference of ion penetration depths. The expression for cumulative charge density distribution, using $S_2(z) = S_1(z) - \Phi(z)$, consists of two parts: the main term and the additional term reflecting ion separation:
\begin{equation}\label{eq:ions_distr_2}
    \rho(z)=-2ec_0 S_1(z) \frac{\sinh\left( \beta e \phi\right)}{1 +2\gamma \sinh^2\left(\beta e \phi/2\right)}+ ec_0 \Phi(z)\frac{ e^{ \beta e \phi}}{1 +2\gamma \sinh^2\left(\beta e \phi/2\right)}
\end{equation}
This equation in dimensionless form with the assumption for Heaviside-like characteristic function $S_1(z)$, which turns to 1 at $z \in [0, +\infty)$, and expressing $\Phi(z) = \mu g(z)$, gives:
\begin{equation}\label{eq:ions_distr_2_dmls}
    \rho(x)=-\frac{\sinh(u)}{1 +2\gamma \sinh^2\left(u/2\right)}+ \frac{0.5\mu g(x)e^{ u}}{1 +2\gamma \sinh^2\left(u/2\right)}
\end{equation}
where $x = z/L_D$, with the Debye length $L_D$, $u = \beta e \phi$ is dimensionless electrostatic potential, $\mu = \Delta/\sigma$ is the magnitude of ion separation, i.e the ratio of difference of penetration depth to surface height deviation, $g(x) = \frac{1}{\sqrt{2\pi}} \exp\{-\frac {x^2}{2\sigma^2}\}$ has the form of the Gauss distribution function. We refer the readers to our previous work [23] if more details are required.

Then, we substitute Eq.\ref{eq:ions_distr_2_dmls} into the Modified Poisson equation from the BSK model [9], and obtain the main equation we solve in the present work:
\begin{equation}\label{eq:problem_appx}
(1-\delta_c^2 \nabla^2) \nabla^2 u = \frac{\sinh(u)}{1 +2\gamma \sinh^2\left(u/2\right)} - \frac{0.5 \mu g(x)e^{u}}{1 +2\gamma \sinh^2\left(u/2\right)}\\
\end{equation}

\subsection{Boundary conditions}

The problem in Eq.\ref{eq:problem_appx} is considered with boundary conditions: $u|_{x=0}= V$, where $V$ is dimensionless electrostatic potential on the electrode and $u'''|_{x=0}= 0$, which reflects electroneutrality near the electrode surface. 
However, the accurate formulation of boundary conditions near rough heterogeneous surface may rise several issues. What is treated as $x=0$ in our model? How to treat electrode surface roughness within simple 1D problem formulation properly?

Obviously, that the structure of electrode surface modifies electrostatic field, and, to accurately describe EDL near rough electrode, it is necessary to consider equipotential boundary condition on rough surface in 2D formulation, like $u(z_s(x,y)) = V$, where $z_s(x,y)$ is the function describing electrode surface structure. To transform this into 1D formulation, one should carry accurate averaging by electrode surface realizations to determine the area affected by the boundary condition on the rough electrode surface and the form of its contribution. Nonetheless, the detailed consideration of boundary conditions on rough electrode seems redundant when the BSK model is applied for short-range electrostatic correlations. We think that the accurate boundary condition should be appreciable with account for hard-sphere interactions in ionic liquids. Even if we accurately account for electrode structure in boundary condition its contribution may be minor compared to the steric effect from electrode roughness. Moreover, the most significant effect from electrode surface structure is observed at low potentials, where the electrostatic effect from rough boundary condition can be neglected. Therefore, we consider these boundary conditions to be appropriate. 

The presented model is most suitable for a flat charged surface with neutral surface modifiers, like precipitated molecules or nanoparticles, and should be considered an assumption for a rough charged surface. In the boundary conditions, the level $x = 0$ corresponds to the mean electrode surface height deviation. Since, we transform the characteristic functions $S_i(z)$ to the Heaviside functions, it means that we consider EDL near charged flat surface, while keeping the additional term describing ion separation near rough surface.

\section{Numerical solution}
\subsection{Details of the numerical solution}

To solve numerically the problem, of describing EDL near rough electrode surfaces, several discretization methods can be applied. We refer the readers to an excellent review of the methods for unbounded domains [25]. In short, there are four groups of methods to approximate a solution on unbounded domains: 

\begin{enumerate}
    \item domain truncation methods with a finite difference or finite element discretization and artificial regularization
    \item approximation by a spectral method with classical orthogonal systems on unbounded domains (e.g. Laguerre or Hermite polynomials)
    \item approximation by spectral method with other, non-classical orthogonal systems, such that the regularity condition is satisfied (e.g. mapped Jacobi polynomials)
    \item mapping of the unbounded domains to bounded domains and application of the standard spectral approximation (with Chebyshev or Lagrange polynomials) to solve the mapped equations in the bounded domains
\end{enumerate}

The first class of methods suffers from low accuracy compared to global (spectral) methods and restrictions on setting boundary conditions at infinity [26,27]. The drawbacks of the second class of methods are related to the calculation of integrals of dot products on infinite domains, the difficulty of setting boundary conditions at 0, and the assumption that the solutions decrease exponentially fast and do not oscillate [28,29]. To apply, the third class of methods, the calculation of dot products on infinite domains is required, and there is the complex procedure of imposing boundary conditions at 0, for example, see Ref.[30].  The fourth class of approximations allows one to apply collocation methods, which do not require the calculation of integrals, the boundary conditions at 0 are imposed easily and they are computationally more efficient. These methods also provide the approximation to the widest class of solutions. Please note, that each of the listed groups of methods can be considered optimal for certain classes of problems. We intend to use the most generic and computationally lightweight method, so we choose the fourth variant.

As mentioned in the main paper, we consider the following quasi-linear fourth-order partial differential equation:
\BEA
\label{supl:main_equation}
\alpha (u)_{xxxx} + \beta (u)_{xx} = f(u,x),\\
u(0) = u_0, (u)_{xxx}(0) = 0,
\EEA
where $\alpha$ and $\beta$ are real numbers, $x \in \Omega:=[0,+\infty]$, $f(u,x)$ is the smooth nonlinear function, and $u(x)$ are solutions of \eqref{supl:main_equation} for which we additionally require regularity at infinity. Further, the notation $({f})_x$ indicates a partial derivative of a function $f$ with respect to $x$. It is also assumed that exists some $x_* \in \Omega: x_* < +\infty$ s.t. $\forall x > x_*$ the frequency of solution oscillations decays at least algebraically and is zero at infinity.

One can show, that the right-hand side function in \eqref{supl:main_equation} is regular at infinity, i.e.:
\begin{equation}
\begin{matrix}
\displaystyle \lim_{x \to \infty}  f(u,x,\sigma,\mu, \gamma) = \frac{\sinh (u)}{1 -\gamma +\gamma \cosh (u)},\\
\displaystyle \lim_{x \to \infty} \frac{\partial^n f(u,x,\sigma,\mu, \gamma)}{\partial^n x} = 0,
\end{matrix}
\end{equation}
with any finite integer $n$, assuming that $u(x)$ is regular at infinity, $\mu \in \mathbb{R}^+$, $\sigma \in \mathbb{R}^+$ and $\gamma \in \mathbb{R}^+$ is such, that $1 -\gamma +\gamma \cosh (u) \neq 0$. 

Using the above-mentioned assumptions one can apply a pseudospectral method to solve the problem \eqref{supl:main_equation} numerically. We selected the set of Chebyshev polynomials in trigonometric form, following a canonical paper [31]:
\[
\phi_j(t) = \cos(j t), t \in[0,\pi],
\]
with the following domain mappings from physical domain $\Omega:=[0,+\infty]$ to basis domain $[0, \pi]$ and back:
\BEA
\label{supl:infty_mapping}
x:=L \cot^2 \left(t/2\right), && t:=2 \text{arccot} \left( \sqrt{x/L} \right),
\EEA
where $L \in \mathbb{R}^{+}$ is a mapping parameter. A particular value of the parameter is selected depending on the problem and is subject to numerical analysis. 

We can note that the selected polynomials are orthogonal on $\Omega$ with a certain weight:
\BEA
\label{supl:orthogonal}
\int_\Omega \phi_j\left(2 \text{arccot} \left( \sqrt{\frac{x}{L}} \right) \right) \phi_k \left(2 \text{arccot} \left( \sqrt{\frac{x}{L}} \right) \right) \left(L \sqrt{\frac{x}{L}}(1+\frac{x}{L})\right)^{-1} dx
=\left\{ 
\BM 
\pi, j=k=0,\\ 
\pi/2, j=k > 0,\\ 
0, j \neq k
\EM \right.
\EEA
Once the mappings \eqref{supl:infty_mapping} and property \eqref{supl:orthogonal} are established, the problem \eqref{supl:main_equation} can be solved in the basis domain. The solution $u$ is expanded in terms of basis functions in the physical and basis domains as follows:
\BEA
\label{supl:basis_expansions}
u(x) = \sum_{j=0}^{\infty} a_j \phi_j\left( 2 \text{arccot} \left( \sqrt{x/L} \right) \right),\\
u(t) = \sum_{j=0}^{\infty} a_j \phi_j(t)
\EEA
where $a_j$ are the expansion coefficients, which are the same for both domains.

The nonlinear system of the pseudospectral (collocation) equations is formed by the substitution of \eqref{supl:basis_expansions} into \eqref{supl:main_equation} and equating the value of the residual to zero at collocation points $t_k, k \in \{0\} \cup \mathbb{N}$ in the basis domain. In this case, one obtains the following infinite system of (generally nonlinear) equations:
\BEA
\label{supl:system_of_nonlinear_equations}
\alpha \sum_{j=0}^{\infty} a_j \left(\phi_j(t_k) \right)_{xxxx} + \beta \sum_{j=0}^{\infty} a_j \left(\phi_j(t_k) \right)_{xx} - f \left(\sum_{j=0}^{\infty} a_j \phi_j(t_k), L \cot \left(t_k/2\right)^2 \right) = 0,\\
u(x)|_{x=0} = u(t)|_{t=\pi} = \sum_{j=0}^{\infty} a_j \phi_j(\pi) = u_0,\\
(u(x))_{xxx}|_{x=0} = (u(t))_{xxx}|_{t=\pi} = \sum_{j=0}^{\infty} a_j (\phi_j(\pi))_{xxx} = 0.
\EEA
To solve the problem with low computational costs, one considers truncated series with $N$ expansion coefficients (number of degrees of freedom or DOF) and ${N-2}$ collocation points. We select collocation points at Chebyshev nodes, i.e:
\BE
\label{supl:collocation_points}
t_k = \pi \frac{2k+1}{2(N-2)}, k = 0,1,...,N-3.
\EE
Thus, we obtain a matrix of size $(N-2) \times N$ that represents the action of the spatial operator on the vector of unknown coefficients. Two additional rows of the matrix are formed by the provided boundary conditions.

The mapping of the derivatives for an arbitrary function $f$ to the basis domain is derived using \eqref{supl:infty_mapping} and chain rule:
\BE
\label{supl:derivative_mapping}
(f)_{x} = \left(d(L \cot \left(t/2\right)^2)/dt\right)^{-1} (f)_{t} = -\frac{1}{2L} \sin^2\left(\frac{t}{2}\right) \tan \left(\frac{t}{2}\right) (f)_{t}.
\EE
Hence, the derivatives $\left(\phi_j(t)\right)_{xx}$ and $\left(\phi_j(t)\right)_{xxxx}$ are constructed by using the mapping \eqref{supl:derivative_mapping}:
\BE
\label{supl:2-nd_derivative}
\left(\phi_j(t)\right)_{xx} =\frac{1}{2 L^2} \sin ^2\left(\frac{t}{2}\right) \tan ^3\left(\frac{t}{2}\right) \left[(\cos (t)+2) (\phi_j(t))_{t} + \sin (t) (\phi_j(t))_{tt}\right]
\EE
\BE
\label{supl:3-d_derivative}
\BM
\left(\phi_j(t)\right)_{xxx} =
-\frac{1}{4 L^3}\sin ^2\left(\frac{t}{2}\right) \tan ^5\left(\frac{t}{2}\right)
\{[6 \cos (t)+\cos (2 t)+8] (\phi_j(t))_{t} +\sin (t)
[ 3 (\cos (t)+2) (\phi_j(t))_{tt} + \sin (t)(\phi_j(t))_{ttt}]\}
\EM
\EE
\BE
\label{supl:4-th_derivative}
\BM
\left(\phi_j(t)\right)_{xxxx} = 
\frac{1}{16 L^4}\sin ^2\left(\frac{t}{2}\right) \tan ^7 \left(\frac{t}{2}\right)\{3 [29 \cos (t)+8 \cos (2 t)+\cos (3 t)+32] (\phi_j(t))_t + \\ \\
 \sin (t) \left[\{72 \cos (t)+11 \cos (2 t)+91\} (\phi_j(t))_{tt}+\right.
\left.2 \sin (t) \{6(\cos (t)+2)(\phi_j(t))_{ttt} + \sin (t)(\phi_j(t))_{tttt}\}\right] \},
\EM
\EE
Note, that for any $\left(\phi_j(t)\right)_{x...}$ one has:
\BE
\lim_{x \rightarrow +\infty} \left( \phi_j \left( 2 \text{arccot} \left( \sqrt{x/L} \right) \right) \right)_{x...}=\lim_{t \rightarrow 0} \left(\phi_j(t) \right)_{x...} = 0,
\EE
and the regularization property at infinity of the solution is automatically satisfied.

\begin{figure}
\centering
\includegraphics[width=1\linewidth]{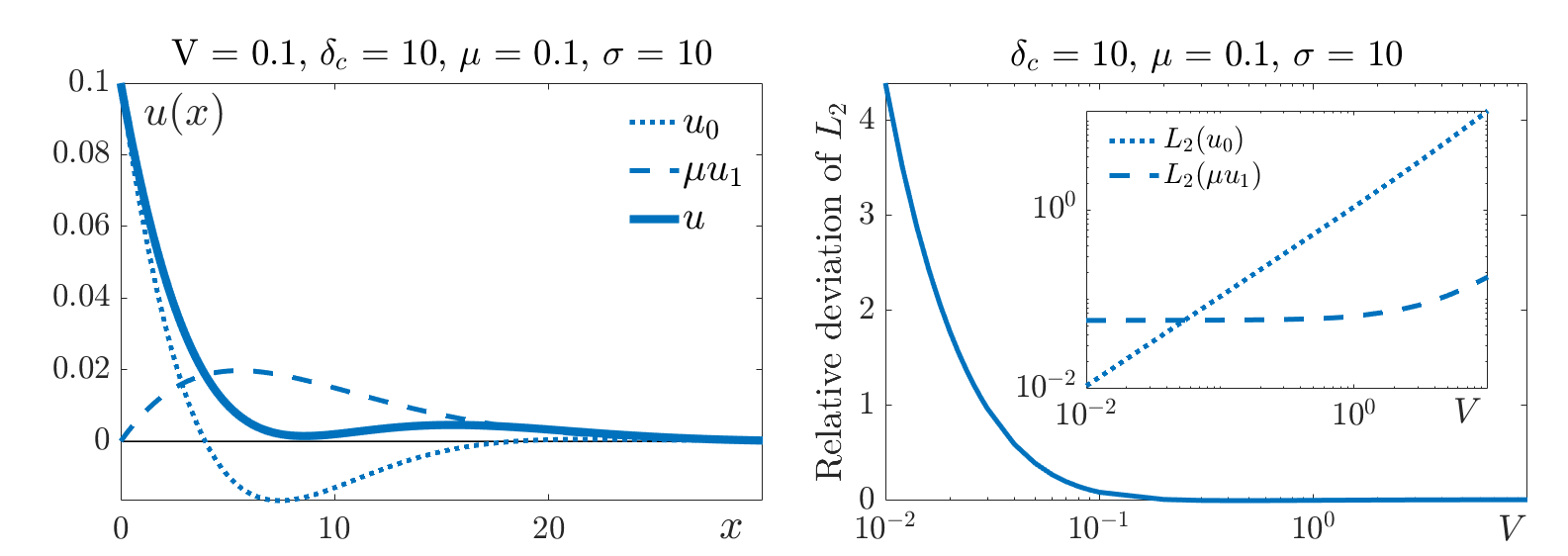}
\caption{On the left: numerical solutions for electrostatic potential near rough electrode at $V = 0.1$, $\sigma = 10$, $\mu = 0.1$, $\gamma = 1$,  $\delta_c = 10$ (the solid line), electrostatic potential near the flat electrode (the pointed line) and additional contribution driven by electrode surface structure (the dashed line). On the right: relative deviation of $L_2$ norm of electrostatic potential near rough electrode from the solution near flat electrode with $V$ at $\mu = 0.1$,  $\sigma = 10$, $\gamma = 1$, $\delta_c = 10$. In the inset, the $L_2$ norm for solution near the flat electrode (the pointed line) and $L_2$ norm of additional term caused by electrode surface structure (the dashed line).}
\label{fig:num_sol}
\end{figure}

Having introduced explicit expressions for derivatives, one needs to impose zero boundary condition on the third derivative at the basis domain. This can be achieved by considering a limit of \eqref{supl:3-d_derivative} at $\pi$ by the L’Hôpital’s rule:
\BE
\lim_{t \rightarrow \pi} \left(\phi_j(t)\right)_{xxx} = -\frac{4 (-1)^j j^2 \left(2 j^4+20 j^2+23\right)}{15 L^3}
\EE

The final system of equations can be written in the form:
\BE
\label{supl:nonliner_basic_system}
A \mathbf{a} = \hat{f}(\mathbf{a}),
\EE
where $A$ is the $N \times N$ non-singular matrix of the spatial differential operator with imposed boundary conditions, $\mathbf{a}$ is the column vector of the unknown expansion coefficients and $\hat{f}$ is the right-hand side function taken at collocation points with two additional values of boundary conditions.

The solution of the nonlinear system of the equations \eqref{supl:nonliner_basic_system} is performed by the Newton-Raphson method [37]. By introducing the Newton iterations one obtains the sequence of solutions $\{\mathbf{a}_{n}\}$ that should converge to the actual solution if the conditions of the convergence theorem are met [32]. Then let $\mathbf{a}_{n+1}=\mathbf{a}_{n}+ \delta \mathbf{a}$. Assume that $\mathbf{a}_0$ is the initial vector of the expansion coefficients and that the function obtained by the substitution of this vector into \eqref{supl:basis_expansions} is regular. Using this approximation in \eqref{supl:nonliner_basic_system} and linearization on the $n$-th step provides the following Jacobi matrix:
\BE
\label{supl:newton_jacobian}
J(\mathbf{a}_n)=J_n:=A - \left.\frac{\partial \hat{f}(\mathbf{a})}{\partial \mathbf{a}}\right|_{\mathbf{a}_n},
\EE
and right-hand side:
\BE
\label{supl:newton_rhs}
g(\mathbf{a}_n)=g_n:=A\mathbf{a}_n - \hat{f}(\mathbf{a}_n).
\EE
Plugging this into the Newton-Raphson iterations one obtains:
\BE
\label{supl:newtons_method}
\left\{
\BM
J_n \delta \mathbf{a} = -g_n,\\
\mathbf{a}_{n+1} = \mathbf{a}_{n} + \gamma \delta \mathbf{a},
\EM\right. n=0,1,...
\EE
where $0<\gamma \leq 1$ is the positive real number ensuring that iterations produce a monotonically decreasing sequence of the right-hand side vectors in \eqref{supl:newtons_method}, i.e.:
\BE
\label{supl:newton_iterations_conditions}
\|g_{n+1}\| < \| g_n\|, \forall n.
\EE

Iterations \eqref{supl:newtons_method} terminate when the right-hand side becomes sufficiently small $\|g_n\| \leq \varepsilon$ for some predefined positive real number $\varepsilon$. If the condition \eqref{supl:newton_iterations_conditions} fails for very small $\gamma$ (close to the machine epsilon), then one needs to perform globalization to obtain the solution. In our case, we apply homotopy globalization (see [33,34]) between smoothed and original nonlinear functions.

In Fig.\ref{fig:num_sol} on the left, we show the results for electrostatic potential profiles near flat and rough electrode surfaces and their difference obtained by our numerical solution. We consider $N = 100$ basis functions, initial solution $\mathbf{a_0}$ is taken from expansion of the function $f(x) =Ve^{-x^2}$ on basis functions and $\varepsilon = 10^{-12}$. We observe that electrode surface structure at the considered parameters makes the electrostatic potential profile positive while keeping its nonmonotonic behavior. In Fig.\ref{fig:num_sol} on the right, we show the relative deviation of $L_2$ norm calculated for the electrostatic potential near the rough electrode from the solution near the flat surface. In the inset, $L_2$ norms of solution near flat and rough electrodes are shown. We can see that there are intersections near the value $V = 0.05$ when the impact of the electrode surface structure becomes less significant than the solution near the flat electrode.

The suggested method is implemented in Python 3 and is available on the repository [36] under Apache license. The algorithm allows one to solve any problem on the semi-unbounded domain, provided that the operators are encoded in the correct form and an arbitrary set of boundary conditions that pose a well-defined problem. It is also assumed that the Jacobi matrix $f(u,x)_u$ and the nonlinear function $f(u,x)$ are provided. The algorithm is implemented using an object-oriented paradigm, which allows one to pass a generic problem class that can be solved by the algorithm. 

\subsection{Validation of numerical solution}

\begin{figure}[!b]
\centering
\includegraphics[width=1\linewidth]{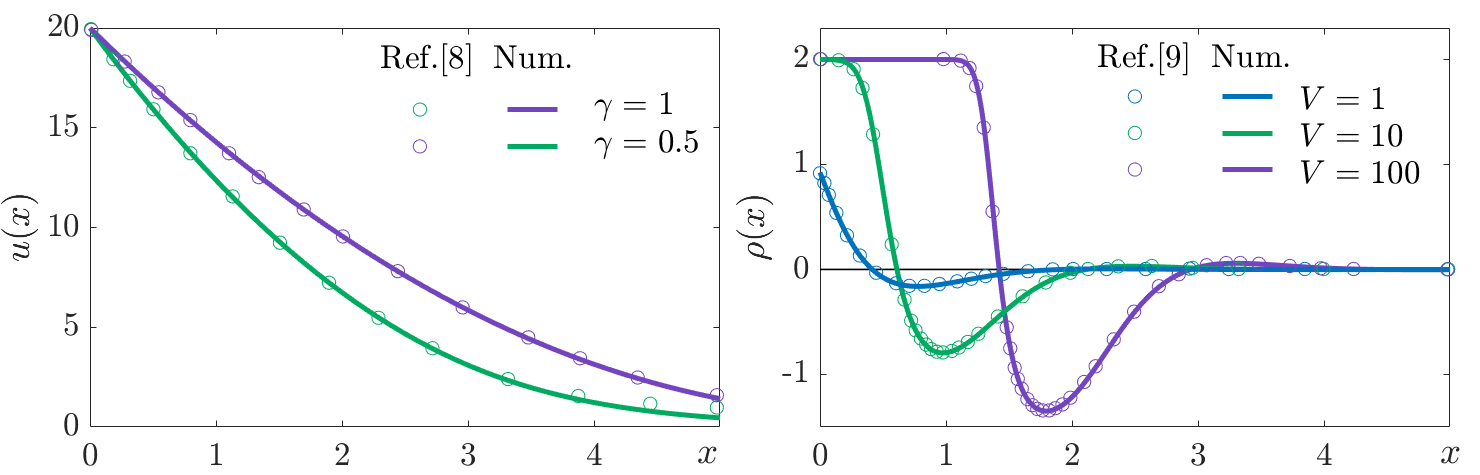}
\caption{On the left: comparison of electrostatic potential profiles from Ref.[8] with solution of our numerical model at $V = 20$ and two values of $\gamma$ = 0.5 and 1. On the right: comparison of charge density from Ref.[9] at various potentials on the electrode $V = 1, 10,$ and 100 and parameters values $\gamma = 0.5$, $\delta_c = 10$.}
\label{fig:validation_BSK}
\end{figure}

To check for the accuracy of our numerical solution, we represent the results reported in the literature [8,9]. In Fig.\ref{fig:validation_BSK} on the left, we present the comparison of the electrostatic potential profiles near flat electrode for two different values of $\gamma$ and find a good quantitative agreement with results from Ref.[8]. Additionally, in Fig.\ref{fig:validation_BSK} on the right, we compare our results for charge density distributions near flat electrode with account for electrostatic correlation at various values of  $V$ on the electrode with the results from Ref.[9]. We achieve very accurate representation, so this validation confirms the reliability and validity of our numerical model.

\section{Analytical solutions}

Next, we solve Eq.\ref{eq:problem_appx} analytically, suggesting $u = u_0 +\mu u_1$ and $\mu \ll 1$. With these suggestions, we can derive the system for the electrostatic potential perturbation $u_1$ given below:
\begin{equation}\label{eq:problem_pert}
\left\{
\begin{array}{lll}
\delta_c^2 \nabla^4 u_1 - \nabla^2 u_1 + u_1 =& 0.5 g(x)e^{u_0}\\
    \left.u_1 \right|_{x=0} &=0&  \\
    \left.u'''_1 \right|_{x=0} &=0&  \\
\end{array}
\right.
\end{equation}
Based on the result obtained by our numerical solver, we distinguish two variants for an analytical solution at $\sigma < \delta_c$ and $\sigma \ge \delta_c$, because the behavior of electrostatic potential perturbation changes in these cases. 

\subsection{Analytical solution for $\sigma < \delta_c$}

In the case of $\sigma < \delta_c$, we use low potential approximation to simplify r.h.s. of the Modified Poisson-Fermi equation in the system \ref{eq:problem_pert} and replace $g(x)$ with a step-like function with the same amplitude $1/\sqrt{2\pi}$ in the area $x \in [0,\tau]$. The system to solve takes the form:
\begin{equation}\label{eq:problem_3_1}
\left\{
\begin{array}{lll}
\delta_c^2 \nabla^4 u_1 - \nabla^2 u_1 + u_1 =& \frac{1}{2\sqrt{2\pi}} (1+u_0)\\
    \left.u_1 \right|_{x=0} &=0&  \\
    \left.u'''_1 \right|_{x=0} &=0&  \\
\end{array}
\right.
\end{equation}

Let's consider the following form for a particular solution:
\begin{equation}
    u_1^p = C e^{-k_1 x} \cos(k_2x) + D e^{-k_1 x} \sin(k_2 x) + E
\end{equation}
where $k_1 = \frac{\sqrt{2\delta_c+1}}{2\delta_c}$, $k_2 = \frac{\sqrt{2\delta_c-1}}{2\delta_c}$, and $C,D,E$ are constants. Substituting it in the equation and considering high values of $\delta_c$, i.e $k_1 = k_2 \approx 1/\sqrt{\delta_c}$ we obtain: $C = -\frac{1}{2} \delta_c V$, $D = \frac{1}{2} \delta_c V$, $E = 1/2$. Combining it with the general solution $u_1^0 = A e^{-k_1 x} \cos(k_2x) + B e^{-k_1 x} \sin(k_2 x)$ and substituting $u_1 = u_1^0 + u_1^p$ into boundary conditions with assumption of high values of $\delta_c$ we obtain:
\begin{equation}\label{eq:solution_3_1}
    u_1 = \frac{1}{2\sqrt{2\pi}}[1 + e^{-kx}(\sin(kx)-\cos(kx))]
\end{equation}

We choose $\tau = 0.5\sigma$ according to validation with numerical solution. For $x>\tau$ (area $II$), we use general solution $u_1^0$ that is joined with solution from Eq.\ref{eq:solution_3_1} (area $I$) in the following way: $u_1^I(\tau) = u_1^{II}(\tau)$ and $du_1^I(\tau)/dx = du_1^{II}(\tau)/dx$. It gives us $A = u_1^I(\tau)$ and $B = A + du_1^I(\tau)/kdx$ for the coefficients of the general solution in the second area. 

This solution has a nonmonotonic behavior with spatial oscillations. Its magnitude grows with the increase of $\sigma$ till achieving the maximum at $x^* = \pi \sqrt{2\delta_c}$. At higher $\sigma$ values, the solution after achieving maximum drops to constant value $\frac{1}{2\sqrt{2\pi}}$, keeping at this value with $x$ increase. However, such behavior does not correlate with the behavior obtained by the numerical solution, so we provide another solution in this case given in the next subsection.  Important to note that, the presented solution does not correlate with potential on the electrode $V$. 

In Fig.\ref{fig:analytical_solutions} on the left,  we show the electrostatic potential profile with the presented analytical solution for perturbation of electrostatic potential at $V = 0.1$, $\delta_c = 10$, $\mu = 0.1$, and $\sigma = 5$. We can see that perturbation changes the potential gradient near the electrode surface, shifts the potential profile on the right, and the well of potential with a negative sign becomes less significant.

\subsection{Analytical solution for $\sigma \ge \delta_c$}

When $\sigma\ge \delta_c$, we take the initial equation of the system \ref{eq:problem_pert}:
\begin{equation}
    \delta_c^2 \nabla^4 u_1 - \nabla^2 u_1 + u_1 = \frac{1}{2} g(x) e^{u_0}
\end{equation}
and change the variable $x$ on $t = x(L_D/\sigma)$, then it takes the following form:
\begin{equation}
    \delta_c^2 (L_D/\sigma)^4 \nabla^4 u_1(t) - \nabla^2 (L_D/\sigma)^2 u_1(t) + u_1(t) = \frac{1}{2} g(t) e^{u_0}
\end{equation}
Accounting for the high values of $\sigma$, we can neglect the fourth derivative, then this equation turns to one we have considered in our previous work [23] and then the solution takes the form:
\begin{equation}
    u_1(x) = \frac{1}{4} \int_0^{\infty} \tilde g(x_0) e^{u_0(x_0)}\left[e^{-|x-x_0|} - e^{-|x+x_0|}\right] d x_0
\end{equation}

This solution on electrostatic potential perturbation is always positive and it has a long-range character. The magnitude of this solution grows with $\sigma$ and the right side of the solution rises providing a long-range response on the electrode surface structure. Besides, in this case, the magnitude of the solution grows with an increase of potential on the electrode $V$. 

\begin{figure}[!h]
\centering
\includegraphics[width=1\linewidth]{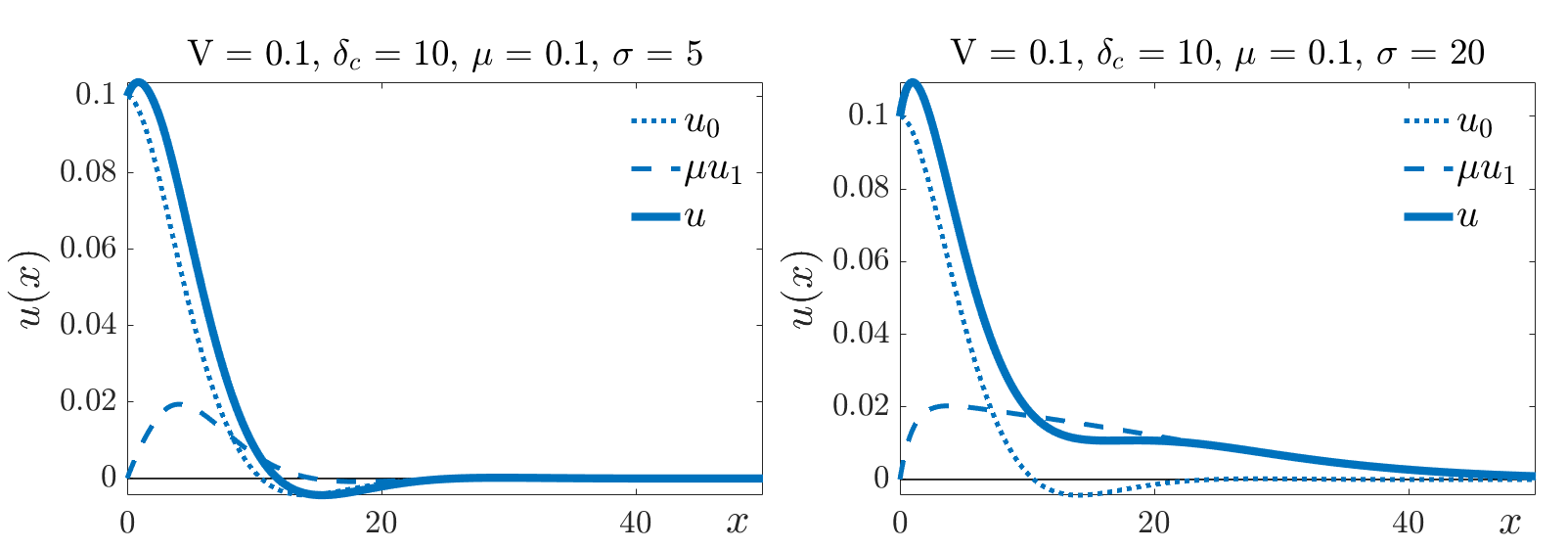}
\caption{Electrostatic potential profiles with account for the electrode surface roughness predicted by our analytical models: on the left, for $\sigma < \delta_c$, and, on the right, for $\sigma \ge \delta_c$. The pointed line shows $u_0$ near the flat electrode from Ref.[9], the dashed line shows the additional term driven by electrode surface structure $\mu u_1$ from our analytical models, and the solid line shows the resulting electrostatic potential $u$, where $u = u_0 + \mu u_1$. We take $V=0.1$, $\delta_c$ = 10, $\mu = 0.1$, and $\sigma$ = 5 for $\sigma < \delta_c$ and $\sigma$ = 20 for $\sigma \ge \delta_c$.}
\label{fig:analytical_solutions}
\end{figure}

In Fig.\ref{fig:analytical_solutions} on the right, we plot the electrostatic potential near the rough electrode surface with this analytical solution on the perturbation of electrostatic potential at the parameters values $V = 0.1$, $\delta_c = 10$, $\mu = 0.1$, and $\sigma = 20$. We observe that electrode surface structure destroys the potential well with negative values, but, it is important to note that it doesn't mean the overscreening breakdown.

\subsection{Analytical solutions transition}
\begin{figure}
\centering
\includegraphics[width=0.5\linewidth]{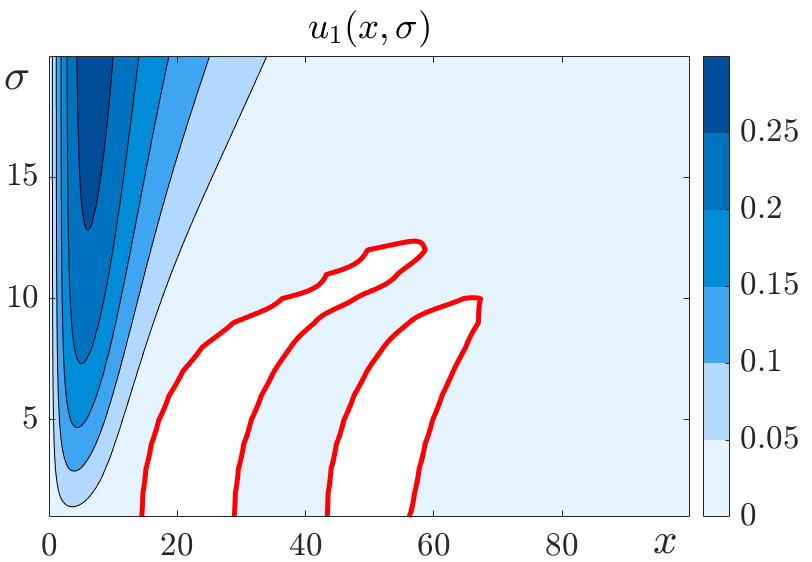}
\caption{Additional term of electrostatic potential driven by electrode roughness $u_1(x, \sigma)$ depending on the value of surface roughness magnitude $\sigma$ calculated by our numerical model. With red line we outline the areas, where $u_1(x, \sigma) < -1 \times 10^{-7}$, to show relatively significant negative values of $u_1(x, \sigma)$.}
\label{fig:solution_transition}
\end{figure}

In this section, we describe the criteria for switching the analytical solutions $\sigma \ge \delta_c$ or $\sigma < \delta_c$. First, we take $\delta_c = 10$ and investigate at which value of surface roughness $\sigma$, the contribution of electrode surface structure to electrostatic potential $u_1 = u - u_0$ calculated by numerical solutions is only positive. We obtain that for higher potential on electrode $V$ a higher value of $\sigma$ is required to make additional electrostatic potential contribution $u_1$ positive. However, since the negative values of $u_1$ are too small, we decided to make this criterion more soft. So, we consider, when $u_1(x) <  -0.01$, to neglect very weak oscillations of $u_1$ sign. For all considered values of potential on electrode $V = 0.1, 0.5,$ and $1$, we obtain that the solutions can be switched at $\sigma \approx 7.3-7.4$. 

In Fig.\ref{fig:solution_transition}, we show  $u_1(x,\sigma)$ at $V = 0.1$ and $\mu = 1$ obtained numerically. With a red line we plot, when $u_1(x) < -10^{-7}$, and observe that nonmonotonic behavior disappear at $\sigma \approx 13$. Actually, $u_1(x)$ has a lot of sign oscillations in the area $x \in [0, \infty)$, we highlight the most significant negative values of $u_1(x)$ with magnitude higher than $10^{-7}$. Since, more accurate estimation for disappearance of nonmonotonic behavior $u_1(x) < -10^{-7}$ is about $\sigma = 13$ and less accurate, where $u_1(x) < -0.01$, is about $\sigma = 7$, we conclude that the idea to switch the solutions at $\sigma = \delta_c$ is correct enough.

\subsection{Validation of numerical and analytical solutions}

\begin{figure}[!h]
\centering
\includegraphics[width=1\linewidth]{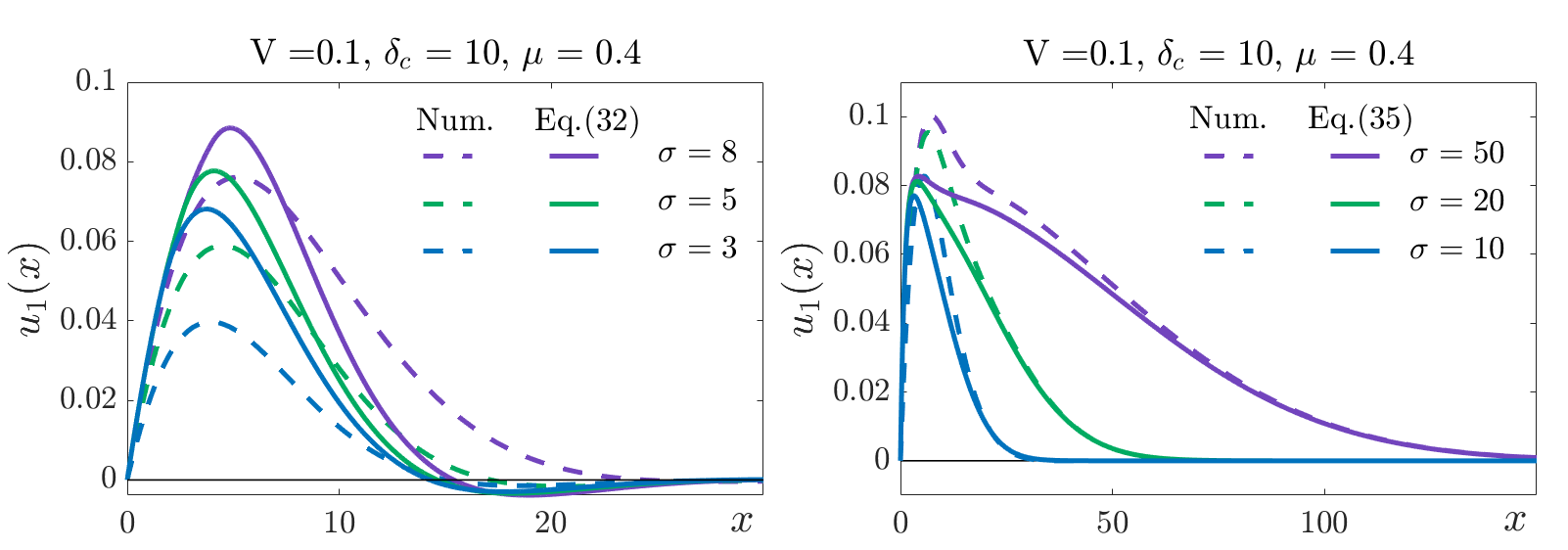}
\caption{Comparison of analytical and numerical solutions for additional term of electrostatic potential driven by electrode roughness at $V = 0.1, \delta_c = 10, \mu = 0.4$, and, on the left: for $\sigma < \delta_c$ with $\sigma = 3, 5$, and $8$, on the right: for $\sigma \ge \delta_c$ with $\sigma = 10, 20$, and $50$.}
\label{fig:comparison}
\end{figure}

To check the validity of analytical solutions, we compare the results for the contribution of electrode surface structure to the electrostatic potential obtained numerically and analytically. We find good qualitative agreement in both cases $\sigma < \delta_c$ (see  Fig.\ref{fig:comparison} on the left) and $\sigma \ge \delta_c$ (see  Fig.\ref{fig:comparison} on the right), while there is some deviation in the magnitude between numerical and analytical solutions. The biggest deviation is obtained in the case $\sigma < \delta_c$, when $\sigma \ll 1$. Overall, the analytical solution for $\sigma < \delta_c$ overestimates the numerical solution, while the analytical solution for $\sigma \ge \delta_c$ underestimates the numerical one.

\newpage

\section{Differential capacitance}

\begin{figure} [!h]
\centering
\includegraphics[width=1\linewidth]{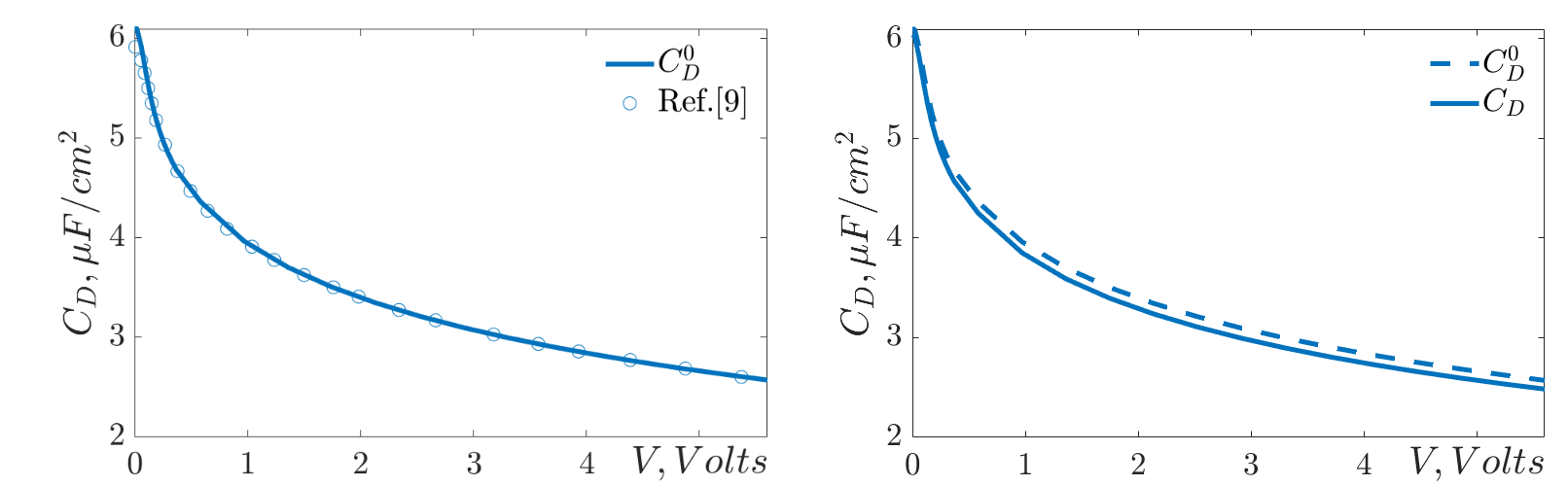}
\caption{On the left: comparison of DC calculated by our numerical model (solid line) near flat electrode with data from Ref.[9] (circles) at $\gamma = 0.5, \epsilon = 5$, and $\delta_c = 10$. On the right: DC near flat electrode at the same parameters (dashed line), and near rough electrode with $\mu = 0.4$, and $\sigma = 10$ (solid line) obtained by our numerical model.}
\label{fig:dc_validation}
\end{figure}

Before studying differential capacitance (DC) modification with electrode surface roughness, first, we compare the results of DC calculation obtained by our numerical model with data from Ref.[9] (see the left part of Fig.\ref{fig:dc_validation}). We take $\delta_c = 10$, $\gamma = 0.5$, $\epsilon = 5$ and used the correction for Stern layer. Here, we multiply the dimensionless differential capacitance $C_D$ on $\epsilon \epsilon_0/L_D$, with  $L_D \approx 0.10361$ nm, and the dimensionless potential on the electrode $V$ on the factor $k_B T/e$ with $T = 450 K$. We can see, that our numerical model manages to accurately represent the data from Ref.[9].

Then, we consider how electrode surface structure changes DC, and apply our numerical model with parameters $\mu = 0.4$, and $\sigma = 10$. In Fig.\ref{fig:dc_validation} on the right, we compare the result for DC profiles near flat $C_D^0$ and rough $C_D$ electrode surfaces. We carry calculations at various values of potential on the electrode $V$ and obtain differential capacitance as a numerical derivative of cumulative charge on potential increment. We can see that the rough electrode provides a small decrease in differential capacitance, which can be related to the worsened ion separation, but this shift is not significant about 2\%.

Next, we move to the analytical estimation for DC perturbation. Since, the analytical solution on electrostatic potential perturbation in the case $\sigma < \delta_c$ does not correlate with $V$, it means that there is no impact on DC in this case. While, when $\sigma \ge \delta_c$, we can obtain an analytical expression for DC perturbation, as follows:
\begin{equation}\label{eq:dc_start}
    C_D^1 = -\frac{\partial}{\partial V}\frac{\partial \mu u_1(x) }{\partial x}= -\frac{\Delta}{2 \sigma} \frac{\partial}{\partial V}\int_0^{\infty} g(x_0) e^{u_0(x_0)}e^{-x_0} d x_0
\end{equation}
Substituting $u_0 = V e^{-k_1 x_0} (\cos(k_2x_0) + A \sin(k_2 x_0))$, with $k_1 = \frac{\sqrt{2\delta_c+1}}{2\delta_c}$, $k_2 = \frac{\sqrt{2\delta_c-1}}{2\delta_c}$, and $A = \frac{\sqrt{2\delta_c+1}}{\sqrt{2\delta_c-1}}\frac{\delta_c-1}{\delta_c+1}$ from Ref.[9] gives:
\begin{equation}\label{eq:dc_analytical}
    C_D^1 = -\frac{1}{2 \sqrt{2\pi}} \frac{\Delta}{\sigma} \int_0^{\infty} e^{-x_0^2/2\sigma^2 -(k_1+1) x_0} (\cos(k_2x_0) + A \sin(k_2 x_0)) e^{u_0(x_0)} d x_0
\end{equation}

If we apply low potential approximation to Eq.\ref{eq:dc_start}, i.e $e^{u_0(x_0)} \approx 1+u_0(x_0)$, we obtain:
\begin{equation}
    C_D^1 \approx-\frac{1}{2 \sqrt{2\pi}} \frac{\Delta}{\sigma} \frac{\partial}{\partial V}\int_0^{\infty} e^{-x_0^2/2\sigma^2}(1+u_0(x_0))e^{-x_0} d x_0
\end{equation}
Substituting $u_0 = V e^{-k_1 x_0} (\cos(k_2x_0) + A \sin(k_2 x_0))$ gives:
\begin{equation} \label{eq:dc_analytical_point}
    C_D^1 \approx-\frac{1}{2 \sqrt{2\pi}} \frac{\Delta}{\sigma}\int_0^{\infty} e^{-x_0^2/2\sigma^2 -(k_1+1) x_0} (\cos(k_2x) + A \sin(k_2 x)) d x_0
\end{equation}
In the limit $\sigma \gg 1$, we can integrate in the following way:
\begin{equation}
    C_D^1 \approx -\frac{1}{2 \sqrt{2\pi}} \frac{\Delta}{\sigma}\frac{k_1 + 1 + A k_2}{(k_1+1)^2+k_2^2}
\end{equation}
and, applying the approximation for $\delta_c \gg 1$, where $A \approx 1$, and $k_1 = k_2 \approx 1/\sqrt{2\delta_c}$, one can obtain:
\begin{equation}
    C_D^1 \approx -\frac{1}{2 \sqrt{2\pi}}\frac{\Delta}{\sigma} \frac{2k + 1}{2k^2+2k + 1} 
\end{equation}
Thus, the electrode surface roughness reduces differential capacitance at low voltages.

\begin{figure}
\centering
\includegraphics[width=1\linewidth]{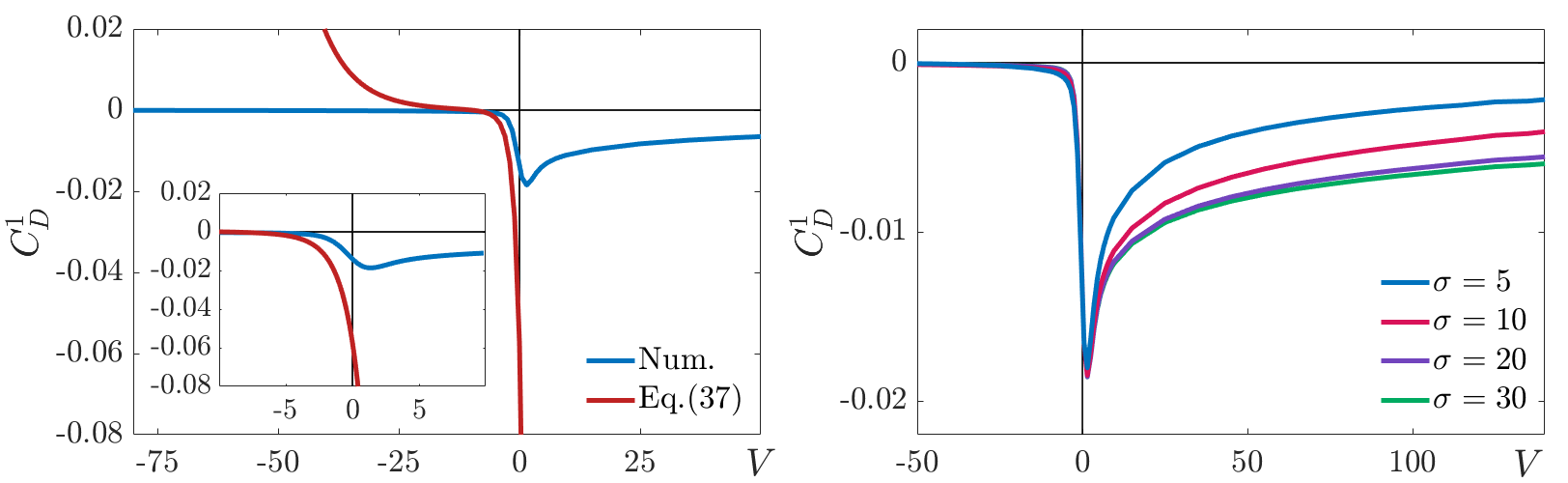}
\caption{On the left: DC decrement driven by electrode surface structure with potential on the electrode obtained numerically (blue line) compared with the result of our analytical estimation from Eq.\ref{eq:dc_analytical} (red line) at $\delta_c = 10$, $\mu = 0.4$, and $\sigma = 10$. On the right:
DC decrement curve with potential on the electrode obtained numerically for different values of the scale of electrode roughness $\sigma = 5, 10, 20, 30$.}
\label{fig:dc}
\end{figure}

In Fig.\ref{fig:dc} on the left, we compare the numerical results for DC increment with our analytical estimation from Eq.\ref{eq:dc_analytical} at $\delta_c = 10$, $\mu = 0.4$, and $\sigma = 10$, using dimensionless values. Our analytical approach allows us to represent DC perturbation near $V = 0$ by employing a low potential approximation form for $u_0$ and assuming high values of $\sigma$, thereby neglecting the fourth-order derivative of the electrostatic potential. Interestingly, the numerical result for the DC shift is not monotonic and approaches zero at high negative potentials. This behavior is attributed to the absence of anions to repel at high negative potentials, as there is a dense cation layer near the electrode surface. The most significant impact of electrode surface structure on DC occurs at small positive values of $V$.

Additionally, we investigate how DC increment varies with the scale of electrode surface roughness $\sigma$. Figure \ref{fig:dc} on the right demonstrates the results of our numerical model for DC increment at $\delta_c = 10, \mu = 0.4$ and various values of $\sigma = 5, 10, 20,$ and $30$. The DC increment decreases for smaller values of $\sigma$. This is consistent with no impact on DC in the analytical solution for $\sigma < \delta_c$. 

Important to note, that this behavior of DC increment with potential on the electrode $V$ is specific to ionic liquids with larger anions. In the case of larger cations, the result will be inverse: there will be no impact at high positive potential, and the most significant alteration of DC will occur at small negative potentials. Probably, if one accounts for specific ion attraction, it could enhance DC. 

\section{Critical potential}

\begin{figure}[h]
\centering
\includegraphics[width=0.8\linewidth]{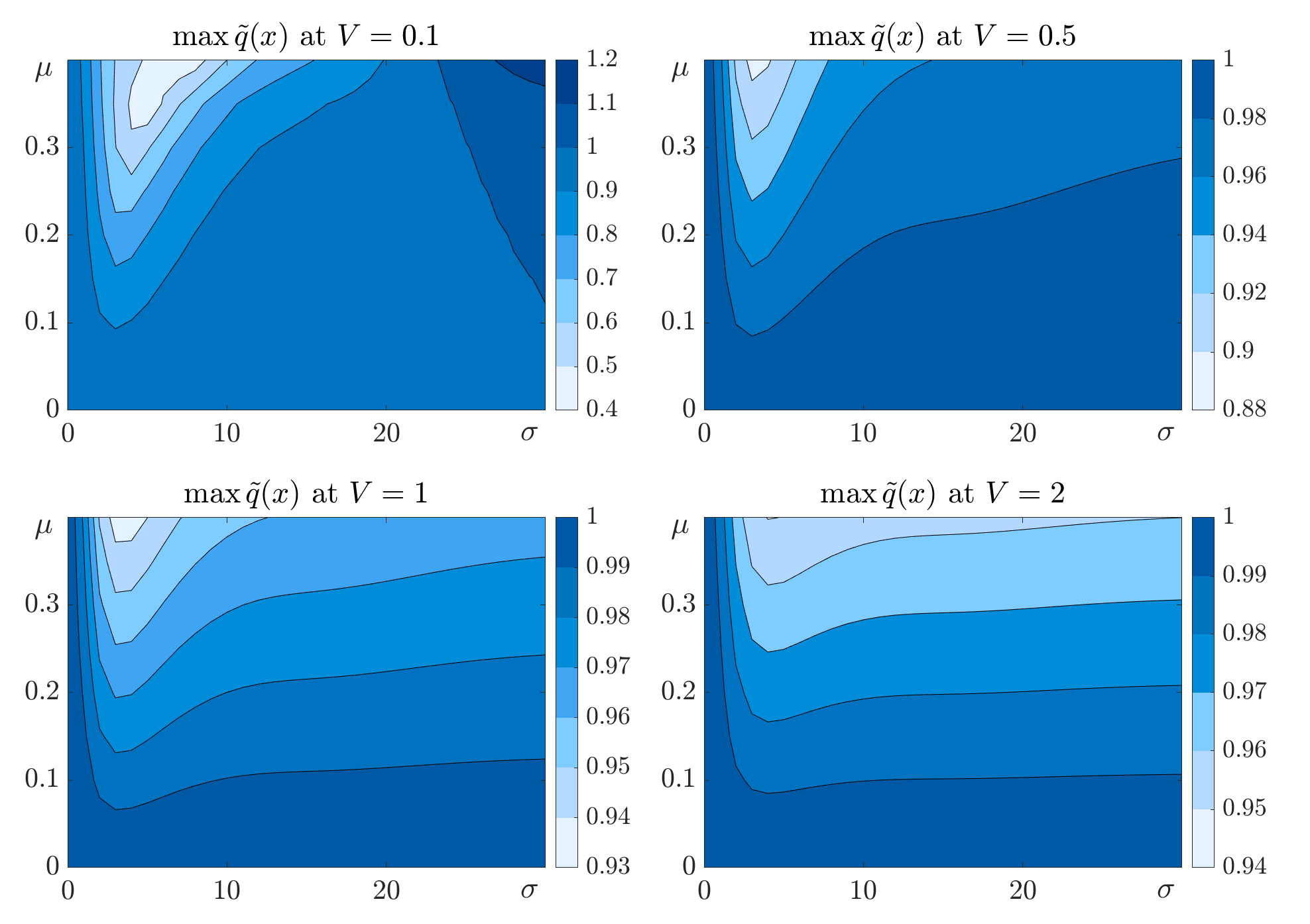}
\caption{Maps of the normed cumulative charge $\tilde q(x)$ magnitude depending on the parameters $\mu \in [0, 0.4]$ and $\sigma \in [0, 30]$ for various potentials on the electrode $V = 0.1, 0.5, 1,$ and $2$.}
\label{fig:q_maps_with_V}
\end{figure}

It is essential to study the constraints when the impact of electrode surface roughness is significant. As we know, the additional term of electrostatic potential driven by electrode surface roughness grows slower than electrostatic potential near a flat electrode with an increase of potential on the electrode (See Fig.\ref{fig:num_sol} on the right). Thus, it is expected that there will be no significant impact on electrode surface structure at high potentials on the electrode. To investigate it, we examine the influence of electrode surface roughness on the EDL structure with potential growth.

First, we consider a magnitude alteration of the normed cumulative charge $\tilde q(x)$ caused by electrode surface structure for different values of potential on the electrode $V$. In Fig.\ref{fig:q_maps_with_V}, we show the maps of the normed cumulative charge magnitude in dependence with parameters $\mu \in [0, 0.4]$ and $\sigma \in [0, 30]$ for various values of the electrostatic potential on the electrode $V = 0.1, 0.5, 1$, and 2. Our results indicate that the impact of electrode roughness on EDL structure becomes less significant with a potential increase.  Additionally, we observe that the area of the most significant drop of $\max \tilde q(x)$ remains consistent with increasing potential.

Next, we take the parameters $\mu = 0.38$ and $\sigma = 6$, which provide the strongest modifications to the normed cumulative charge at $V = 0.1$ and study for layering alterations with potential increase. To analyze it, we consider charge densities of EDL near flat and rough electrode surfaces, distinguish the layers by the sign of charge density, and utilize Shannon—Wiener entropy calculations to measure systems structural similarity. This metric is highly sensitive to layering modifications because we reinterpret it as binary functions that can only be shifted and scaled in one dimension, making it particularly sensitive to objects overlap statistics [35]. Specifically, it indicates how frequently the layers coincide; when layers overlap to the maximum extent, entropy becomes close to 0. 

\begin{figure}
\centering
\includegraphics[width=1\linewidth]{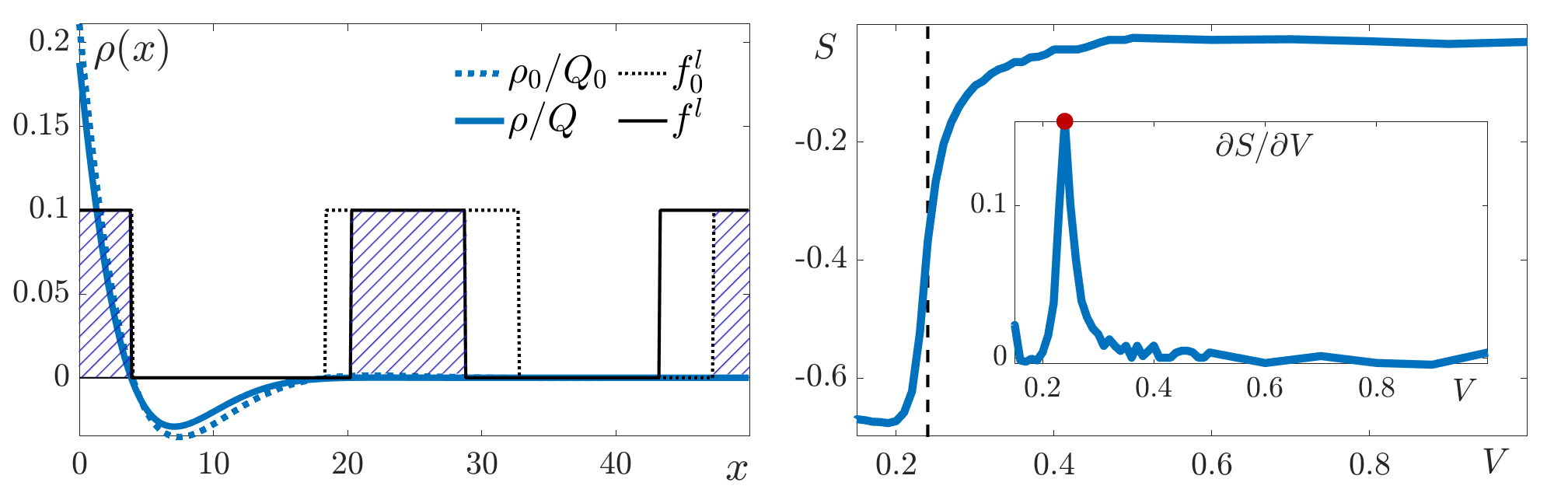}
\caption{On the left: profiles of the normed charge density and layering functions for EDL near flat and rough electrodes at $V = 0.24, \delta_c = 10, \mu = 0.38$, and $\sigma = 6$. Here, the layering function $f^l$ is scaled on the value 0.1 for representativity. On the right: the entropy-potential curve and its derivative (in the inset), showing the structure similarity of the EDL near flat and rough surfaces at $\delta_c = 10, \mu = 0.38$ and $\sigma = 6$.}
\label{fig:entropy}
\end{figure}

Let us consider charge density $\rho(x)$ and the layering function $f^l(x)$:
\begin{equation}
    f^l(x) = \left\{
    \begin{matrix}
        0, &\text{if $\rho(x)/Q < 0$}\\
        1,  &\text{if $\rho(x)/Q \ge 0$}\\
    \end{matrix}\right.
\end{equation}
where $Q = \int_0^{\infty} \rho(x) dx$. The function $f^l(x)$ indicates the layers in the structure and its sign with the sign of the cumulative charge of EDL. When we consider positive potentials on the electrode, then the cumulative charge of the EDL is negative, and $f^l(x)$ will turn to 1, when it distinguishes a layer with a greater number of anions, and 0 in the opposite case. In the left part of Fig.\ref{fig:entropy}, we show the charge densities normed on the cumulative charge and the corresponding layering functions for EDL near flat and rough electrode surfaces. The layering function is scaled to a value of 0.1 for representativity, and dashed lines highlight the areas where layering structures coincide, contributing to entropy evaluation.

Then, we calculate the probabilities of layers coincidence of EDLs near flat and rough surfaces and their Shannon—Wiener entropy:
\begin{equation}
    p_{11} = \int f^l_0(x) f^l(x) dx, \quad p_{00} = \int \bar f^l_0(x) \bar f^l(x) dx, \quad S = \sum_{i = {0, 1}} p_{ii}\log p_{ii}
\end{equation}
where $f^l_0(x)$ is the layering function near flat electrode and $f^l(x)$ is the layering function for EDL near rough electrode.

In Fig.\ref{fig:entropy} on the right, we plot the entropy-potential curve considering the parameters $\mu = 0.38$ and $\sigma = 6$. One can see that at low potential the entropy is minimal, which means quite different structures of EDL near flat and rough electrode surfaces. When the potential on the electrode increases, the entropy approaches 0, causing the EDL structures to become similar. We take the value of the electrode potential at which we observe the maximum of the entropy gradient $V^* = 0.24$ as the critical potential when roughness stops to modify the layering structure significantly.

\begin{figure}
\centering
\includegraphics[width=1\linewidth]{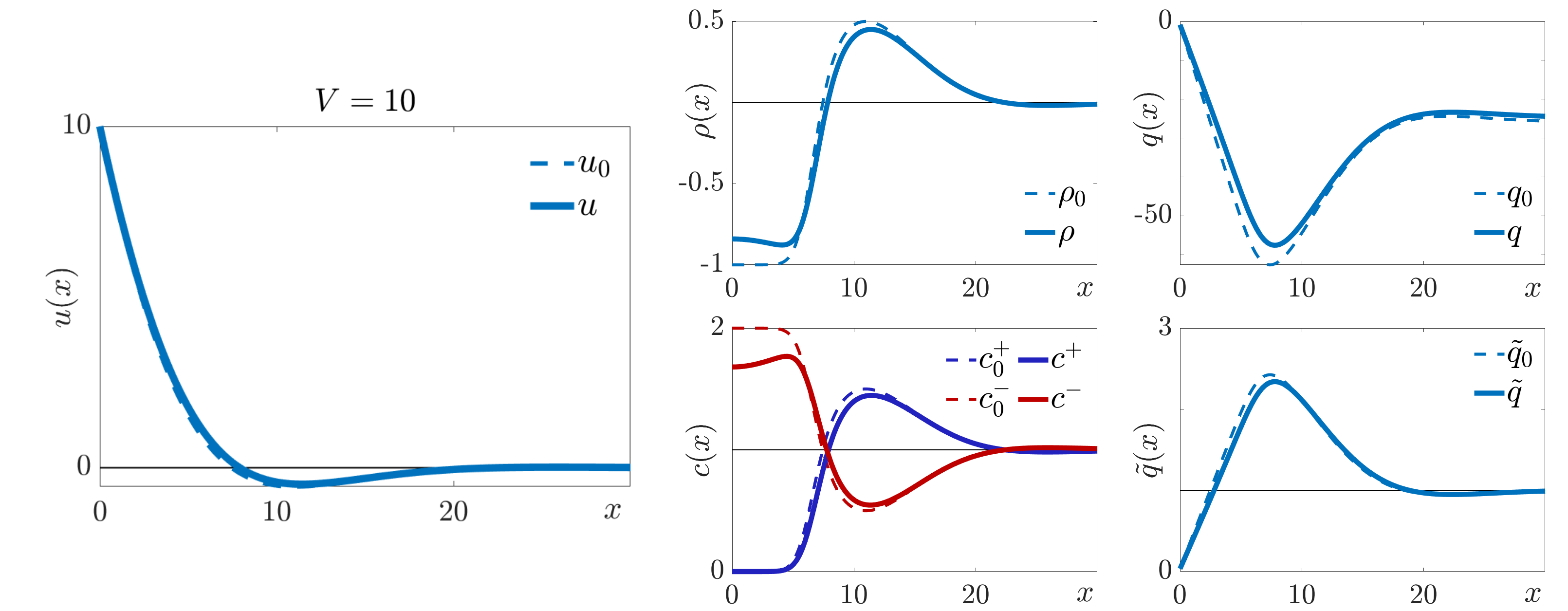}
\caption{Comparison of electrostatic potential $u(x)$, charge density $\rho(x)$, ion concentrations $c_{\pm}(x)$, cumulative charge $q(x)$, and normed cumulative charge $\tilde q(x)$  near rough (solid lines) and flat (dashed lines) electrode surfaces at $V = 10, \delta_c = 10, \mu = 0.4$, and $\sigma = 5$.}
\label{fig:crowding}
\end{figure}

It is also interesting to check how the EDL near the rough electrode is changed at high electrode potential, so we consider the crowding regime. In Fig.\ref{fig:crowding}, we show the electrostatic potential $u(x)$, charge density $\rho(x)$, ion concentrations $c_{\pm}(x)$, cumulative charge of electric double layer $q(x)$, and the normed cumulative charge profile $\tilde q(x) = \int_0^x \rho(x)/\int_0^{\infty} \rho(x)$ for the case of crowding regime at $V = 10$. We observe that the charge density in the area near the electrode surface and the following overscreening become less significant. It occurs due to the small repulsion of anions from the electrode surface. The cumulative charge rises a bit, especially in the area where crowding turns to overscreening, which leads to a little decrease in the normed cumulative charge. To sum up, one can see that there are no significant modifications in the EDL structure and properties driven by electrode surface roughness at high positive potentials when the crowding regime takes place.

\begin{figure}
\centering
\includegraphics[width=0.5\linewidth]{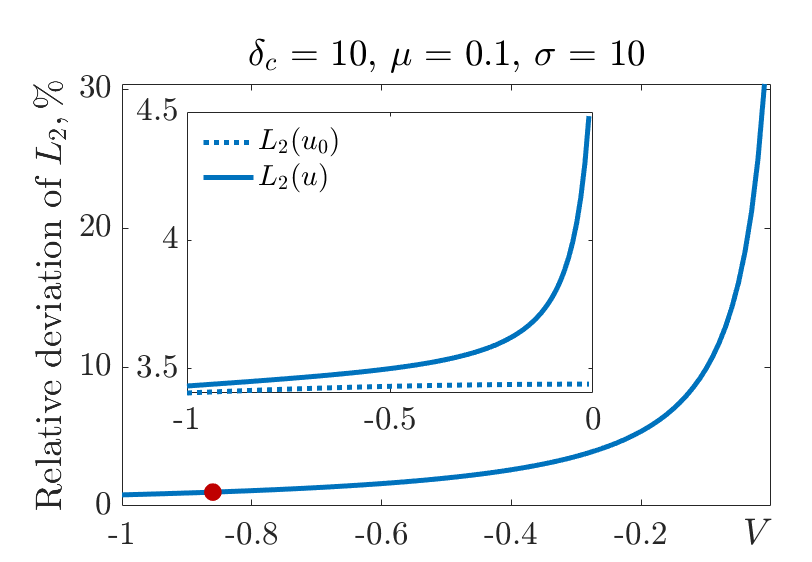}
\caption{Relative deviation of $L_2$ norm of numerical solution for electrostatic potential near rough electrode from the solution near flat electrode for $V<0$ at $\mu = 0.4$,  $\sigma = 2$, $\gamma = 1$, $\delta_c = 10$. In the inset, the $L_2$ norm for the solution near the flat electrode (the pointed line) and $L_2$ solution for the solution near the rough electrode (the solid line)}
\label{fig:critical_V_neg}
\end{figure}

Finally, we estimated critical potential for $V<0$, while as the are no structural modifications, we apply the criteria on $L_2$ norm alteration. In Fig.\ref{fig:critical_V_neg}, we plot the deviation of $L_2$ norm calculated for electrostatic potential near rough and flat electrode surfaces, which are illustrated in the inset. With a red point, we highlight the value $V^* = -0.86$, which corresponds to $L_2$ norm deviation 1\%. Thus, if consider $V<-0.86$, we can neglect the impact of electrode surface structure, as anions would be repelled from the surface by electrostatic forces.

\end{document}